\documentclass[a4paper,11pt]{article}
\pdfoutput=1 

\usepackage{jheppub} 
\usepackage{bbm,bm,graphicx,mathtools,color,hyperref,slashed}
\usepackage{amssymb,amsmath}
\usepackage{comment}
\usepackage[all]{xy}
\usepackage[dvipsnames]{xcolor}

\usepackage{preamble}

\graphicspath{{./Figures/}}

\newcommand{\diff}{\mathrm{d}}

\newcommand{\im}{\mathrm{i}}

\newcommand{\calA}{\mathcal{A}}

\newcommand{\rme}{\mathrm{e}}

\def\Z{{\mathbb Z}}
\def\R{{\mathbb R}}
\preprint{}

\def\coeff#1#2{{\textstyle {\frac {#1}{#2}}}}

\def\half{\coeff 12}

\title{Refined instanton analysis of the 2D $\CP^{N-1}$ model: mass gap, theta dependence, and mirror symmetry}

\author{Mendel Nguyen, Mithat \"{U}nsal, }

\affiliation{Department of Physics, North Carolina State University, Raleigh, NC 27607, USA}

\emailAdd{mendelnguyen@gmail.com, unsal.mithat@gmail.com}

\abstract{
We address nonperturbative dynamics of the two-dimensional bosonic and supersymmetric $\CP^{N-1}$ models for general $N$ by developing new tools directly on $\RR^2$. 
The  analysis starts with a new formulation of instantons that is consistent with the existence of the classical moduli space, classical dipole--dipole type interactions of instanton--anti-instanton pairs, and vanishing interaction of instanton--instanton pairs. 
The classical consistency is achieved via a representation of the instanton as a collection of $N$ pointlike constituents carrying pair of real and imaginary charges valued in the weight lattice of $\SU(N)$.
The constituents interact via a generalized  Coulomb interaction and do not violate the fact that instanton is a single lump with integer topological charge. 
By developing the appropriate Gibbs distribution, we show that the vacuum can be captured by a statistical field theory of these constituents, and their cluster expansion.  
Contrary to the common belief that instantons do not capture the vacuum structure and non-perturbative properties of such theories, our refined analysis is able to produce properties such as mass gap, theta dependence, and confinement of the theory on $\RR^2$. 
In supersymmetric theory, our construction gives a new derivation of the mirror symmetry between the sigma model and the dual Landau--Ginzburg model by Hori and Vafa.  
Our construction also demonstrates that there is absolutely no conflict between large $N$ and instantons.}

\begin{document}

\maketitle

\section{Introduction}

Instantons were originally hoped to provide a mechanism for dynamical mass generation and color confinement in four-dimensional gauge theories \cite{Polyakov:1987ez}. 
So far, this hope has not yet been realized. 
A major source of difficulty is the fact that, for classically scale-invariant theories, instantons can be of any size, and integrating over this size leads to an infrared divergence in the dilute gas approximation. 
Moreover, it has been argued that instanton effects are irrelevant in the large $N$ limit \cite{Witten:1978bc}. 
Thus, to the extent that large $N$ analysis is trusted, instantons do not seem to play a decisive role in determining key features of the dynamics.

The purpose of this paper is to challenge this pessimism.
We wish to argue that a refined treatment of instantons \emph{can} in fact provide a reliable account of the nonperturbative physics, in sharp contrast with the history of this subject. 
We do so by working in the context of the two-dimensional $\CP^{N-1}$ model, where similar problems with the instantons of four-dimensional gauge theory are present.

Our treatment relies on the following ingredients: 
\begin{itemize}
    \item[(1)] a description of the instanton moduli space in terms of the positions of ``pseudoconstituents'' of an instanton;
    \item[(2)] a proper treatment of classical instanton--instanton and instanton--anti-instanton interactions; 
    \item[(3)] an appropriate Gibbs distribution for the grand canonical ensemble of pseudoconstituents; 
    \item [(4)] the inclusion of subextensive contributions to the cluster expansion.
\end{itemize} 
As a result of our analysis, we are able to correctly capture important qualitative features of the dynamics such as the existence of a mass gap and the multibranched structure of the $\theta$ dependence. 
Applied to the $\mathcal{N}=(2,2)$ supersymmetric version of the model, we are able to reproduce the mirror  symmetry results of Hori and Vafa \cite{Hori:2000kt} by using nonsupersymmetric tools.

Now let us say a bit more about each of these ingredients:

\paragraph{(1) Pseudoconstituents of an instanton.} 

It is an old idea that the existence of ``fractional instantons'' would avoid the problems which have beset the instantons. 
For instance, the argument about the irrelevance of instantons in large $N$ would not apply to fractional instantons. 
Indeed, the claim that instantons should be irrelevant in large $N$ follows from the fact that their effects are weighted by $e^{-c/g^2}$, with $c$ a positive constant independent of $g$ and $N$. 
Since the large $N$ limit is obtained by keeping $N g^2$ finite, instantons appear to be exponentially suppressed as $N \to \infty$. 
On the other hand, the fractional instantons have action $1/N$ times that of ordinary instantons, and hence their effects would be weighted instead by $e^{-c/N g^2}$, which does not vanish as $N \to \infty$ (with $Ng^2$ fixed).

The attractiveness of the fractional instanton idea has led many authors in the past to look for them as constituents of the ordinary instanton on $\RR^2$ \cite{Gross:1977wu, Berg:1979uq, Fateev:1979dc, Bukhvostov:1980sn}. 
\footnote{There are also circumstances  in which instanton physically fractionates, either by compactification on 
 $\RR^1 \times \SS^1$ with flavor twisted boundary conditions, or on $\RR^2$ with the introduction of some extra fields. In these cases,  the instanton physically splits up  into $N$ fractional instantons 
\cite{Dunne:2012ae,Unsal:2020yeh,Brendel:2009mp,Eto:2004rz,Tong:2002hi}, which are themselves  are physical configurations. 
Both procedures comes with the cost of introducing an extra dimensionful scale in the problem.}
It is known that the moduli space of unit topological charge instantons has dimension $2N$ (more generally, the moduli space of topological charge $q$ instantons has dimension $2 N q$), and this is compatible with the idea that an instanton should be made up of $N$ pointlike constituents, each contributing two moduli corresponding to its position in spacetime. 
Another clue comes from the classical interactions between instantons and/or anti-instantons (about which we will say more below).
As found by Gross and others \cite{Gross:1977wu, Callan:1977gz}, for an instanton--anti-instanton pair, the interaction is of dipole--dipole type. 
Taking the picture of the instanton as a dipole literally, one is led to speculate that the monopole constituents of which the dipole is presumably made should somehow correspond to fractional instantons.

The approach of this paper, while having some superficial similarities, departs from the fractional instanton program in a crucial way. 
We parameterize the $2N$-dimensional instanton moduli space in terms of $N$ complex coordinates $a_i$, and interpret each $a_i$ as the spacetime position of some kind of object.
However, these are \emph{not} fractional instantons -- we make no attempt to identify these objects with any fractional action solution of the classical Euclidean field equations.
In fact, we make no attempt to identify them with any particular field configuration at all.
Only the collection $a_1,\ldots,a_N$ can be associated with a single lump, the ordinary instanton. 
For this reason, we refer to these objects as ``pseudoconstituents'' of the instanton (although for brevity, we will often just say ``constituents'').  
For us, what is important is not whether these constituents can be associated with particular field configurations, but that they can be used as an effective description of the instanton gas in such a way that the classical instanton--instanton and instanton--anti-instanton interactions are correctly captured, and the structure of the classical moduli space is respected.

\paragraph{(2) Classical interactions between instantons and/or anti-instantons.} 

Since the classical instanton--anti-instanton interaction is of dipole--dipole type, we should like to view each constituent as some kind of charged object by analogy with two-dimensional electrostatics \cite{Gross:1977wu}.
However, the analogy between instantons and electrostatic dipoles cannot be the whole truth. 
This is because, crucially, the classical instanton--instanton interaction is exactly zero, as is implied by the very existence of a multi-instanton moduli space. 
Thus, any assignment of real charges to the constituents, while it may correctly reproduce the instanton--anti-instanton interaction, cannot correctly reproduce the vanishing instanton--instanton interaction. 

For the case of $N=2$, it actually has been proposed by Forster \cite{Forster:1977jv} that it is possible to assign charges to the constituents consistent with the classical interactions if each constituent is assigned two kinds of charges, one real and one imaginary charge.\footnote{However, the statistical field theory constructed in Ref.\cite{Forster:1977jv} is incorrect. Perhaps this is the reason this interesting idea did not receive much attention.} 
As a result, instantons acquire two kinds of dipole moments, one real and one imaginary.  
In fact, the choice of the constituent charges is such that one has the pair of dipole moments $(\Vec{d},+i\Vec{d}) $ for the instanton and $(\Vec{d}^{\,\prime},-i\Vec{d}^{\,\prime}) $ for the anti-instanton, where $\Vec{d}, \Vec{d}^{\,\prime}$ are the spacetime 2-vector respectively describing the relative displacement of the instanton constituents or the anti-instanton constituents, so that the instanton--instanton and instanton--anti-instanton interactions are respectively
\begin{align}
    U_{\inst \inst} &\propto (1,+i) \cdot (1,+i) = 0 \\
    U_{\smash{\inst \antiinst}} &\propto (1,+i) \cdot (1,-i) \neq 0
\end{align}
as required. 

In our point of view, any analysis based on instantons must correctly account for the classical interactions.  
We will show that, for general $\CP^{N-1}$, a correct description of the classical interactions between instantons and/or anti-instantons is achieved by assigning each constituent both a real $(N-1)$-dimensional charge and an imaginary $(N-1)$-dimensional charge. 
In fact, the $i$-th constituent of the instanton/anti-instanton is given the charges $(\bm{\nu}_i, \pm i \bm {\nu}_i )$, where $\bm{\nu}_i$ is the $i$-th weight of the fundamental representation of $\SU(N)$. 

\begin{figure}[t]
\vspace{-1.5cm}
\begin{center}
\includegraphics[width = 0.9\textwidth]{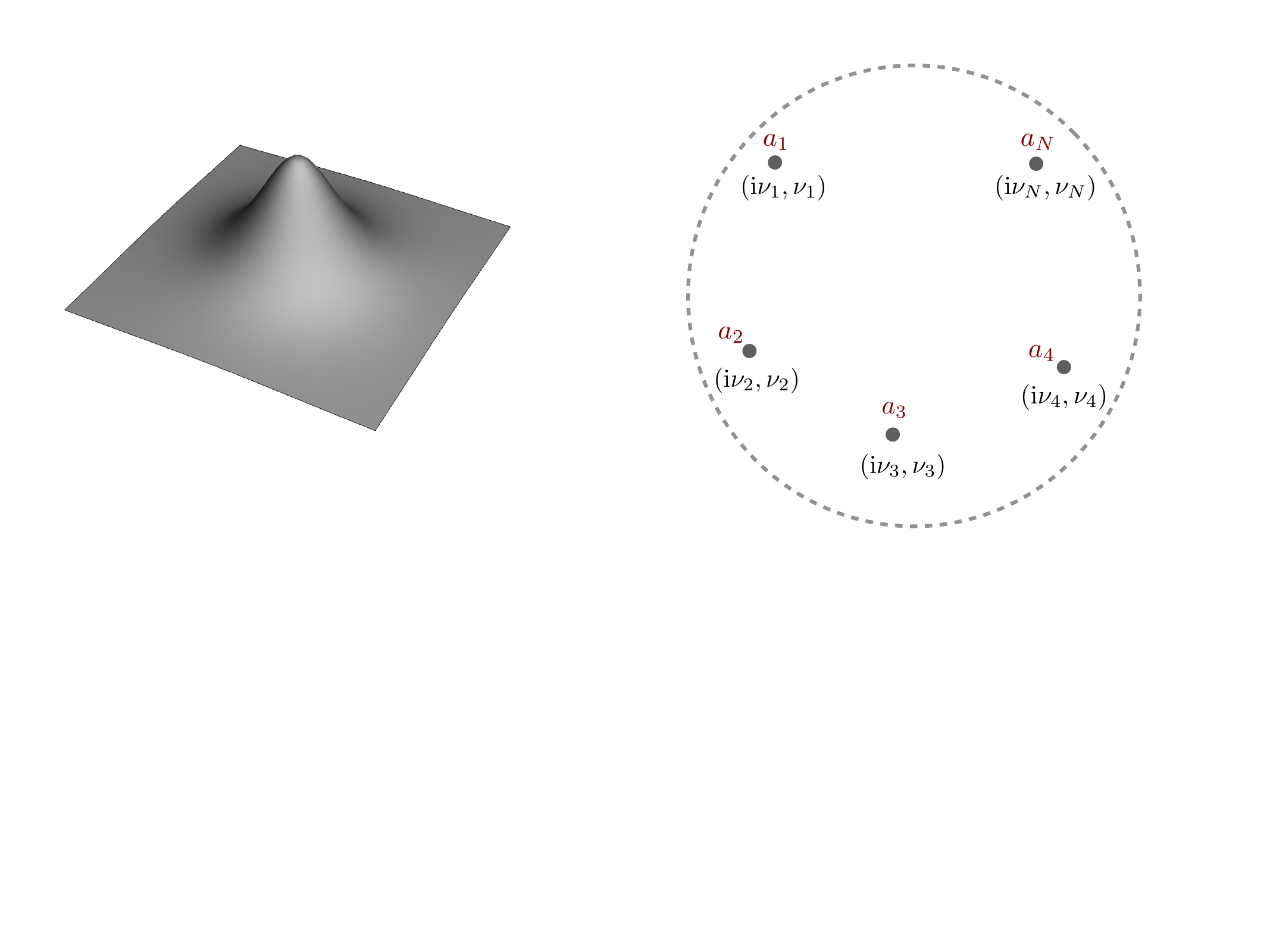}
\vspace{-4.5cm}
\caption{Left: Topological charge density of a single  instanton on $\mathbb R^2$ is a single lump.  Clearly, there is no fractionalization. Right: $a_i, i=1, \ldots, N$ denotes 
$2N$ moduli parameters. Capturing the classical instanton-antiinstanton interaction and absence thereof for an instanton-instanton pair  require assigning a combination of real and imaginary charges to the moduli parameters.  The charges are valued in a real and imaginary copy of  
weight lattice as 
$(i {\bm \nu}_i,   {\bm \nu}_i ) \in   i \weightlattice \oplus \weightlattice $  of $\SU(N)$.
 }
\label{fig:FwF}
\end{center}
\end{figure}

\paragraph{(3) Gibbs distribution.}

Let us consider the $\CP^{N-1} $  model in some very large volume $V$, where  instantons are described in terms of their constituents as in Fig.\ref{fig:FwF}.   
In a grand canonical ensemble, or equivalently, Gibbs distribution with a variable number of particles, we consider the whole system and the subsystem.  
Denote the subsystem as $V_{\text{sub}}$ and environment as the complement $V \backslash  V_{\text{sub}}$.  
A subsystem in thermal equilibrium with its surrounding can exchange both energy and particles with the environment.  
The number of particles (for us, instanton constituents) in the subsystem will necessarily vary.  
Denote by $(n_1, \ldots, n_N)$ and  $(\bar n_1, \ldots, \bar n_N), $ the number of constituents of types   
 \begin{align}
 (i {\bm \nu}_i,   {\bm \nu}_i ), \qquad  (i {\bm \nu}_i,   - {\bm \nu}_i ) \qquad (i=1,\ldots,N)
 \end{align}
in the subsystem  $V_{\text{sub}}$. 
Our main observations about the subsystem are the following:
\begin{itemize}
    \item Variation of a moduli parameter will necessarily take any constituent out of the subsystem $V_{\text{sub}}$, or  will let in some constituents that were out,  regardless of where the instanton is originally  located.  
    \item For the purposes of modeling the instanton gas by a statistical field theory, it is as if we have an ensemble of  classical point particles with charges    $(i {\bm \nu}_i,   {\bm \nu}_i )$. 
    The number of constituents inside subsystem $V_{\text{sub}}$ with charges  $( \pm i {\bm \nu}_i,   {\bm \nu}_i )$  will necessarily fluctuate about their mean value. 
\end{itemize}
Let us stress the crucial difference between, on the one hand, theories in which instantons have a fixed (or maximum) size, such as spontaneously broken  gauge theories, and on the other, theories with classical scale invariance, in which instantons come in all sizes. 
For systems with finite instanton size, there are of course some instantons which lie partly inside and partly outside the subsystem. 
However, since the instantons have a maximum size, these are resticted to appear in the vicinity of subsystem boundary. 
It follows that their contribution to physical observables (in the thermodynamic limit) is of measure zero.  
For classically scale invariant theories, however, variation of moduli parameters will take part of the instanton inside the subsystem, even if the instanton center is held fixed far outside.
Therefore, for the former class of theories, the Gibbs ensemble can be restricted to standard instantons, while for the latter, the Gibbs ensemble must be based on pseudoconstituents as described above. This is an essential difference between preexisting works (such as \cite{Witten:1978bc, Coleman198802}) and ours.

\begin{figure}[t]
\vspace{-1.5cm}
\begin{center}
\includegraphics[width = 1.05\textwidth]{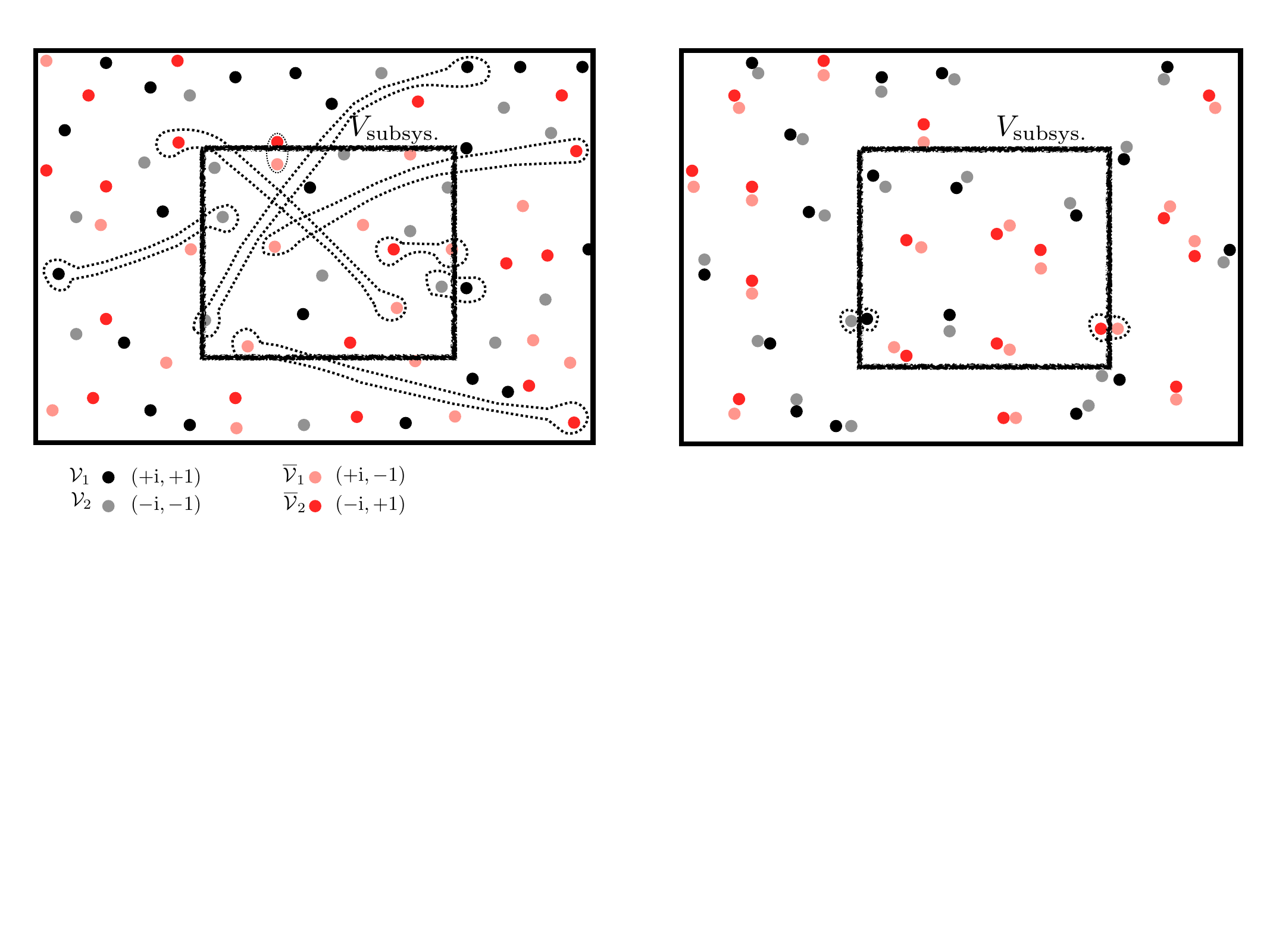}
\vspace{-6.0 cm}\caption{
Left: In classically scale invariant theories, instantons can be of any size at the classical level.   Regardless of the position of the center of an instanton, as the moduli parameters are varied, all constituents will independently enter and exit the subsystem.  
Right: For theories with fixed instanton size,  a constituent of an instanton can  enter or exit the subsystem if and only if  instanton is in sufficiently close vicinity of the subsystem boundary. Therefore, the Gibbs ensemble for classically scale invariant theories must be based on constituents.  For theories with fixed instanton size, it is consistent to build a Gibbs ensemble on standard instantons. This is because, unlike in the case of classically scale invariant theories, the contribution of boundary straddling instantons to vacuum properties are of measure zero in the thermodynamic limit.}
\label{fig:plasma1}
\end{center}
\end{figure}

Next, we construct a  statistical  field theory based on the proliferation of constituents. 
The operator corresponding to a constituent is   
\begin{align}
{\cal V}_i  \sim 
e^{-y_i} \equiv e^{-\varrho_i + i \vartheta_i} \qquad (i=1, \ldots, N)
\label{vorop}
\end{align}
where $y_i = \varrho_i - i \vartheta_i$ is a complex scalar field dual to the original sigma model field. 
A priori, there is no topological charge or action attached to these operators.  
Our only request is that the product of these $N$ operators gives us an instanton, 
i.e., $\prod_{i=1}^{N} e^{-y_i} \sim  e^{-I + i \theta}$. 
Based on this, a dual Lagrangian can be written. 
The  critical point of the dual formulation is the mean field for $y_i(x)$ field. 
There are actually $N$ possible values for the mean field 
\begin{align}
\frac{\delta S_{\rm eff}}{\delta y_i} =0 \Longrightarrow \langle e^{-y_i(x)} \rangle =e^{ - \frac{I}{N} + i \frac{(\theta + 2 \pi k)}{N}}. 
\end{align}
This means, at the mean field level, all constituents have the same complex fugacity (or density.)

This, in our opinion, is an interesting situation. 
Despite the fact that there is no a priori fractional action associated with the constituents, at the level of the mean field, they behave as if they have action $S= \frac{I}{N}$ and topological charge $\frac{1}{N}$. 
Yet, there is no physical lump in the  Euclidean description with fractional action or topological charge!   

In the supersymmetric theory, the mean field values correspond to $N$ isolated vacua of the theory, associated with the spontaneous breakdown of discrete chiral symmetry. 
In the bosonic theory, we obtain the same mean field values, but now they account for the multiple branches non-degenerate quasivacua.  
This shows that the multibranched theta dependence can emerge from the instanton analysis.  
The mean field is stable in the supersymmetric theory and unstable in the bosonic theory. 
However, we show that the second order effects in fugacity expansion can stabilize the mean field. 
This brings us to the cluster expansion.

\paragraph{(4) Cluster expansion.}

The last important ingredient of our construction is the cluster expansion. 
In the supersymmetric theory, the leading order vortex operators are accompanied by fermionic zero modes, and the second order effects in fugacity are usually calculated from the superpotential.  
Here, we evaluate them  via the cluster expansion. 
The evaluation of two-cluster contributions requires the use of Lefschetz thimbles for the quasi-zero mode directions along the lines of \cite{Behtash:2018voa}. 
The only two-cluster contributions are
 \begin{align} 
\left.
\begin{array} {ll}
 {\cal B}_{ii} = [ {\cal V}_i \overline {\cal V}_i ]  :  & \qquad  (i 2  \nu_i,  0) \cr
{\cal B}_{ij}  =   [ {\cal V}_i \overline {\cal V}_j ]  : &  \qquad  (i (\nu_i + \nu_j),  (\nu_i - \nu_j)  ) \cr 
\end{array} \right\} \in  i  \weightlattice \oplus \weightlattice 
\label{list-cluster-2}
\end{align}
This coincides precisely with the bosonic potential derived from the superpotential of the supersymmetric theory. 
As a result, this construction gives a new derivation of the mirror symmetry between a supersymmetric gauged linear sigma model (GLSM) and a
Landau--Ginzburg (LG) dual description of Toda type.

\section{Classical aspects of \texorpdfstring{$\CP^{N-1}$}{CPN-1} model instantons}

This section reviews some known facts about the $\CP^{N-1}$ model instantons, giving special attention to the dipole--dipole form of the instanton--anti-instanton interaction. While this dipole--dipole interaction has been derived before, our presentation will have some new elements which will be useful for our analysis later. 

Let us begin by recalling two well-known formulations of the $\CP^{N-1}$ sigma model \cite{DAdda:1978vbw, DAdda:1978dle}. 
One is what we shall refer to as the gauged formulation. 
The fields are $N$ complex scalars $\bm \phi = (\phi_1,\ldots,\phi_N)$, subject to the constraint
\begin{align}
    \bm{\phibar} \cdot \bm{\phi} = \sum_{i=1}^N |\phi_i|^2 = 1
\end{align}
as well as a gauge equivalence under spacetime-dependent phase rotations
\begin{align}
    \bm \phi(x) \mapsto e^{i\alpha(x)} \bm \phi(x). 
\end{align}
The Euclidean space action is taken to be
\begin{align}
    S = \frac{1}{g^2} \int d^2 x \,|\del_\mu \bm \phi + i A_\mu \bm \phi |^2,
    \label{eq:gauged1}
\end{align}
where $A_\mu$ is the vector field required to implement the gauge principle. Note that since $A_\mu$ has no kinetic term, it may be trivially integrated out by making the substitution $A_\mu = i \bm \phi^\dag \del_\mu \bm \phi$. A topological theta term may also be added:
\begin{align}
    - i \theta \int \frac{dA}{2\pi}. 
\end{align}

The other formulation, which we shall call the geometrical formulation, deals with $N-1$ unconstrained complex scalar fields $w^1,\ldots,w^{N-1}$, and the action is taken to be 
\begin{align}
    S = \frac{1}{g^2} \int d^2x \, G_{r \sbar}(w,\wbar) \,\del_\mu w^r \del_\mu \wbar^{\sbar}
\end{align}
where $G_{r \sbar}$ is the Fubini--Study metric on $\CP^{N-1}$:
\begin{align}
    G_{r \sbar}(w,\wbar) \, dw^r d\wbar^\sbar 
    = \frac{|d\wbar_r dw^r|^2}{1 + \wbar_r w^r} - \frac{|\wbar_r d w^r|^2}{(1 + \wbar_r w^r)^2}.
\end{align}
The relation between the two formulations is given by
\begin{align}
    w_r = \phi_r / \phi_N.
\end{align}

We note that in our normalization of $g^2$, the instanton action is given by
\begin{align}
    I = 2 \pi / g^2,
\end{align}
and thus the one-loop running coupling is given by the RG equation
\begin{align}
    \mu \frac{d}{d\mu} \biggl(\frac{2\pi}{g^2} \biggr) = N.
\end{align}

\subsection{General instanton solutions}

We now review the instanton solutions, pointing out two useful and equivalent parametrizations thereof. 
It will be convenient to work with complex spacetime coordinates $z = x_1 + i x_2, \zbar = x_1 - i x_2$. 
The general instanton solution is given by 
\begin{align}
    \phi_i (z, \zbar ) &= \frac{  \rho u_i + (z-\ac) v_i}{ |  \rho \bm u + (z-\ac) \bm v |}, 
    \label{eq:instsol}
\end{align} 
where $\ac$ is a complex number, $\rho$ is a positive real number, and $\bm u$ and $\bm v$ are orthonormal complex $N$-vectors ($\bm \ubar \cdot \bm u = \bm \vbar \cdot \bm v=1, \bm \ubar \cdot \bm v=0$).
The general anti-instanton solution is obtained by complex conjugating the right-hand side of Eq.~\eqref{eq:instsol}.

The meaning of each of the various parameters is as follows:
$\ac$ and $\rho$ specify the center of action and the size of the instanton, respectively.
For a fixed $\bm v$, the possible values of $\bm u$ make up the intersection of the ($2N-2$)-dimensional subspace orthogonal to $\bm v$ and the unit ($2N-1$)-sphere in $\CC^N$, and this intersection is simply a unit ($2N-3$)-sphere.
The parameter $\bm v$, however, is not a collective coordinate -- it specifies the boundary condition of the field at spacetime infinity:
 \begin{align}
\bm \phi  (|z|  \to \infty) = \frac{z}{|z|} \bm v.
\end{align}
Thus, the instanton moduli space is given by
\begin{align}
    {\cal M}_{N,k=1} = \underbrace{\CC_{}}_{\ac} \times \underbrace{\RR_{>0}}_{\rho} \times \underbrace{\SS^{2N-3}_{}}_{\bm u} = \RR^2 \times (\RR^{2N-2} - {\bm 0}).
    \label{conventional}
\end{align}

Let us now discuss another parametrization the instanton solution. 
This will give a more useful alternative perspective on the moduli space, and help give a physical interpretation of the classical instanton interactions.
Let us show that for a specific choice of boundary condition at spacetime infinity, the instanton solution \eqref{eq:instsol} can be rewritten as
\begin{align}
    \phi_i (z, \zbar) &= \frac{z-a_i}{(\sum_{j=1}^N |z-a_j|^2)^{\frac{1}{2}}}, 
    \label{eq:instsol'}
\end{align}
where $a_1, \ldots, a_N$ are some complex numbers that are only required to not all coincide. It is easy to see that the set of all such $\bm a = (a_1, \ldots, a_N) \in \CC^N$ gives rise to the same moduli space \eqref{conventional}. To make the connection with the form \eqref{eq:instsol} explicit, define parameters $\ac, \rho, \bm u$ by
\begin{align} 
    \ac &= 
    \frac{1}{N} \sum_{i=1}^N a_i , \\
    \rho^2
    &=  
    \frac{1}{N} \sum_{i=1}^N |   \delta a_i   |^2
    =  
    \frac{1}{2N^2} \sum_{i,j=1}^{N} |a_i -a_j|^2 \qquad (\delta a_i \equiv a_i - \ac), \\
    N^{\frac{1}{2}} \rho  u_i 
    &=  
    - \delta a_i.
\label{mapping0}
\end{align}
Using these definitions, the configuration \eqref{eq:instsol'} can be rewritten as
\begin{align}
    \phi_i  (z, \zbar) = N^{-\frac{1}{2}}  \frac{(z -a_{\rm c}) - \delta a_i}{(|z-a_{\rm c}|^2   + \rho^2)^{\frac{1}{2}}},
\end{align}
which is indeed of the form \eqref{eq:instsol} with $\bm v = N^{-\frac{1}{2}}(1,\ldots,1)$. 

It is worth noting that the action density of the instanton depends only the center position $a_{\rm c}$ and the size $\rho$:
\begin{align}
    {\cal L} \propto \frac{ \rho^2} { (|z-a_{\rm c}|^2   + \rho^2)^2 }
\label{profile}
\end{align}
In particular, the parameters $\delta a_i$ ($i=1, \ldots, N$) do \emph{not} enter into the action (or topological charge) density -- the instanton is unequivocally a single lump.
Nevertheless, the $\delta a_i$ do enter in a crucial way into the classical instanton--anti-instanton interaction, which implies, as we shall see below, that one can view an instanton or anti-instanton as a collection of rather unusual pointlike charges.

\subsection{Classical instanton interactions}

To derive the classical instanton interactions, we shall follow the strategy employed by Vainshtein et al. \cite{Vainshtein:1981wh} for the analogous problem in four-dimensional gauge theory. 
The idea is to first find operators, $\cal I$, $\overline{\cal I}$, respectively describing the influence of small instantons or small anti-instantons on large wavelength vacuum fluctuations in the effective theory in which wavelengths larger than these instantons or anti-instantons have been integrated out. Then the instanton interactions may be read off from the two point functions of these  in this effective theory according to
\begin{align}
    \langle {\cal I}(x) {\cal I}(y) \rangle \propto \exp( - V_{\cal I \cal I}(x-y)), \\
    \langle {\cal I}(x) {\overline{\cal I}}(y) \rangle \propto \exp( - V_{\cal I \overline{\cal I}}(x-y)).
\end{align}
Because we are only interested in the classical interactions, the calculation is actually quite simple -- we will only need to work at the tree level. 

It will be most convenient to work in the geometrical formulation, where the general instanton solution \eqref{eq:instsol'} takes the form 
\begin{align}
    w_r(z,\zbar) = \frac{z - a_r}{z - a_N}.
\end{align}
Since we shall have to make contact with perturbation theory, we must work with fields that vanish at spacetime infinity. In our chosen boundary condition, the fields $w_r$ approach the value $1$ at infinity, so let us instead work with the fields
\begin{align}
    \fluc_r \equiv (w_r - 1)/g,
\end{align}
which do vanish at infinity. 
In terms of these new fields, the quadratic part of the action is given by
\begin{align}
    S_* = \frac{1}{N} \sum_{rs} \int d^2 x \, K_{rs}\, \del_\mu \fluc_r \del_\mu \flucbar_s = \frac{4\pi}{N} \sum_{rs} \int \frac{|d\zbar \, dz|}{2\pi} K_{rs}\, \del \fluc_r \delbar \flucbar_s,
    \label{eq:gaussianaction}
\end{align}
where $\del \equiv \del/\del z, \delbar \equiv \del / \del \zbar$, and the matrix $K_{rs}$ is given by
\begin{align}
    K_{rs} = \delta_{rs} - \frac{1}{N}. 
\end{align}
It will be important for us to note that this matrix has a special group theoretic significance -- it is an $(N-1)\times(N-1)$ submatrix of the $N \times N$ matrix of inner products between the weights $\bm \nu_i$ of the defining representation of $\SU(N)$:
\begin{align}
    K_{rs} = \bm \nu_r \cdot \bm \nu_s
\end{align}
(with an appropriate normalization of the $\bm \nu_i$).

We can now proceed to find the operator ${\cal I}_{\delta \bm a}(\ac, \abar_{\rm c})$ corresponding to the instanton configuration specified by the collective coordinates $\bm a = (a_1 ,\ldots,a_N)$. It is requ ired that
\begin{align}
    \Bigl\langle {\cal I}_{\delta \bm a}(\ac,\abar_{\text c}) \prod_{k=1}^{m} \fluc_{p_k}(z_k,\zbar_k) \prod_{l=1}^n \flucbar_{q_l} (z'_l,\zbar'_l) \Bigr\rangle 
    \approx 
    d\mu(\rho) \prod_{k=1}^{m} f_{p_k}(z_k;\bm a) \prod_{l=1}^n \overline{f}_{q_l}(\zbar'_l;\bm \abar)
    \label{eq:matching}
\end{align}
for $|z_k-\ac|, |z_l'-\ac| \gg \rho$, where $d\mu(\rho) \propto (M\rho)^{N}\rho^{-3}d\rho$ is the instanton density and $ f_p(z;\bm a)$ is the value of $\fluc_p$ for the instanton:
\begin{align}
    f_p(z;\bm a)
    \equiv g^{-1}\frac{a_N - a_p}{z-a_N}.
\end{align}
We claim that the operator
\begin{align}
    {\cal I}_{\delta \bm a} = d\mu(\rho)  \exp \biggl( \frac{4\pi}{Ng} \sum_{rs} K_{rs} (a_N - a_r) \del \flucbar_s + \text{c.c.} \biggr)
    \label{eq:inst-op}
\end{align}
does the trick.
To see this, substitute this into the left-hand side of Eq.~\eqref{eq:matching} and expand the exponential. 
The only contributions in the leading semiclassical approximation come from the terms in the expansion with $m$ factors of $\del \flucbar$ and $n$ factors of $\delbar \fluc$, and evaluating with respect to the Gaussian action $S_*$ \eqref{eq:gaussianaction}.
The result is a product of two-point functions:
\begin{align}
    d\mu(\rho) &\prod_{k} \frac{4\pi}{Ng} \sum_{rs} K_{rs} (a_N - a_r) \Bigl\langle \fluc_{p_k}(z_k,\zbar_k) \del \flucbar_s(a_{\text c},\abar_{\text c}) \Bigr\rangle_* \nonumber\\
    &\times \prod_{l} \frac{4\pi}{Ng} \sum_{rs} K_{rs} (\abar_N - \abar_r) \Bigl\langle \flucbar_{q_l}(z_l',\zbar_l') \delbar \fluc_s(a_{\text c},\abar_{\text c}) \Bigr\rangle_*,
\end{align}
where the subscript $*$ indicates a correlator being evaluated with respect to the Gaussian action $S_*$. 
(Here, the operator \eqref{eq:inst-op} is understood to be normal ordered.)
We thus see that in order for Eq.~\eqref{eq:matching} to hold, we must have
\begin{align}
    \frac{4\pi}{Ng} \sum_{rs} K_{rs} (a_N - a_r) \Bigl\langle \fluc_p(z,\zbar) \del \flucbar_s(a_{\text c},\abar_{\text c}) \Bigr\rangle_*
    = f_p(z;\bm a) \approx g^{-1}\frac{a_N - a_p}{z - a_{\text c}}.
\end{align}
This relation is easily checked with the help of the free propagator
\begin{align}
    \langle \fluc_p (z,\zbar) \flucbar_s (z',\zbar') \rangle_* 
    = - \frac{N}{4\pi} (K^{-1})_{ps} \log |z-z'|^2 + \text{const.}
\end{align}
Hence, \eqref{eq:inst-op} is indeed the correct effective instanton vertex. 

In a completely analogous way, we can also obtain the effective vertex for an anti-instanton configuration $\fluc_p = \overline{f_p(z; \bm b)}$: 
\begin{align}
    \overline{\cal I}_{\delta \bm b}= d\mu(\rho)  \exp \biggl( \frac{4\pi}{Ng} \sum_{rs} K_{rs}(b_N - b_r) \del \fluc_s + \text{c.c.} \biggr). 
\end{align}

Let us now compute the classical instanton interactions. From the instanton effective vertex, we can immediately see that instantons cannot interact classically with each other: 
\begin{align}
    U_{\inst \inst} = 0. 
\end{align}
Indeed, the exponent of the instanton operator vanishes if we substitute for $\fluc_s$ any holomorphic function, and a fortiori, any instanton configuration $f_s$. On the other hand, the classical instanton--anti-instanton interaction does not vanish. As noted above, it is extracted by taking the logarithm of the two-point function $\langle \cal I \overline{\cal I} \rangle$. Since in our approximation the instanton operators are exponentials of free fields, the logarithm of the two-point function of the exponentials is simply the two-point function of the exponents; that is,  
\begin{align}
    U_{\cal I \overline{\cal I}}
    &= - \biggl\langle \biggl( \frac{4\pi}{Ng} \sum_{rs} K_{rs} (a_N - a_r) \del \flucbar_s + \text{c.c.} \biggr)
    \biggl( \frac{4\pi}{Ng} \sum_{pq} K_{pq}(b_N - b_p) \del \fluc_q + \text{c.c.} \biggr) \biggr \rangle_* \nonumber\\
    &= \frac{4\pi}{Ng^2(\ac-\bc)^2}\sum_{rp} K_{rp}(a_{N}-a_{r})(b_{N}-b_{p}) + \text{c.c.}
\end{align}
Recalling that $K_{rs} = \bm \nu_r \cdot \bm \nu_s$ and using the relation $\bm \nu_N = - \sum_r \bm \nu_r$, we can further transform this into the following very elegant form:
\begin{align}
    U_{\cal I \overline{\cal I}}
    = 
    -\frac{4\pi}{Ng^2} \frac{(\sum_i \bm \nu_i a_{i}) \cdot ( -\sum_j \bm \nu_j b_{j} )}{(\ac - \bc)^2}  + \text{c.c.}
    \label{eq:dipole-dipole}
\end{align} 
This is the dipole--dipole interaction between an instanton and an anti-instanton.

\section{The refined instanton gas}

The two essential points reviewed in the previous section regarding classical instanton interactions are as follows: first, the interaction between an instanton and an anti-instanton is of dipole--dipole type, and second, the interaction between two instantons vanishes exactly. Taking into account these crucial classical properties will allow us to construct a refinement of the instanton gas, which is the goal of this section. 

\subsection{The necessity of real and imaginary charges}
The formula we have obtained for the dipole--dipole interaction for an instanton--anti-instanton pair, Eq. \eqref{eq:dipole-dipole}, makes it very tempting to view each instanton or anti-instanton collective coordinate as giving the position of a pointlike constituent carrying some kind of multicomponent charge. 
In particular, it would appear that we can simply assign to each instanton collective coordinate $a_i$ the charge $\bm \nu_i \in \weightlattice$, and to each anti-instanton collective coordinate $b_i$ the charge $-\bm \nu_i \in \weightlattice$. 
This would be fine insofar as the instanton--anti-instanton interaction is concerned, but as we have already mentioned in the Introduction, we would not correctly capture the vanishing instanton--instanton interaction. 

This does not mean we have to give up on the idea of viewing the instanton as a collection of charged constituents.
There is in fact a way to do it, which is as peculiar as it is elegant. 
In order to respect both the instanton--instanton and instanton--anti-instanton interactions, we shall assign to each collective coordinate a pair of multicomponent charges, one real and one imaginary, such that the real and imaginary charges do not couple to each other. 
In fact, a charge assignment that works (and which appears to be essentially unique) is given by
\begin{align}
    a_i &\leadsto (+\bm \nu_i, i \bm \nu_i) \in \weightlattice \oplus i \weightlattice, \\
    b_i &\leadsto (-\bm \nu_i, i \bm \nu_i) \in \weightlattice \oplus i \weightlattice;
\end{align}
that is, the $i$-th instanton constituent is assigned the charge $(+\bm \nu_i, i \bm \nu_i)$ and the $i$-th anti-instanton constituent the charge $(-\bm \nu_i, i \bm \nu_i)$. 
In this way, we can achieve the vanishing of the instanton--instanton interaction while maintaining the dipole--dipole interaction for the instanton--anti-instanton pair. 

To see this, let us normalize the 2D Coulomb potential according to
\begin{align}
    G(z,\zbar) = - \frac{2 \pi}{N g^2} \log |z|^2.
\end{align}
Then the interaction between a type $i$ instanton constituent at $a_i$ and a type $j$ anti-instanton constituent at $b_i$ is given by 
\begin{align}
    U_{i j}
    &\equiv (+\bm \nu_i, i \bm \nu_i) \cdot (-\bm \nu_j, i \bm \nu_j) \, G(a_i - b_j, \abar_i - \bbar_j) 
    = 
    \frac{4 \pi}{N g^2} (-\bm \nu_i \cdot \bm \nu_j)(- \log|a_i - b_j|^2).
    \label{eq:constituent-interaction}
\end{align}
Summing over $i,j=1,\ldots,N$ and expanding up to the second order in $\delta a_i, \delta b_j$, we recover the interaction Eq. \eqref{eq:dipole-dipole}:
\begin{align}
    \sum_{i,j=1}^N U_{ij} = \sum_{i,j=1}^{N} \frac{4 \pi}{N g^2} (-\bm \nu_i \cdot \bm \nu_j)(- \log|a_i - b_j|^2)
    \approx 
    -\frac{4\pi}{Ng^2} \frac{(\sum_i \bm \nu_i a_{i}) \cdot ( -\sum_j \bm \nu_j b_{j} )}{(\ac - \bc)^2}  + \text{c.c.}
\end{align}
At the same time, the interaction between two instanton pseudoconstituents at $a_i$ and $a_j'$ is proportional to
\begin{align}
    (+\bm \nu_i, i \bm \nu_i) \cdot (+\bm \nu_j, i \bm \nu_j) = 0,
\end{align}
and hence vanishes. 

\subsection{First steps towards the refined instanton gas}

Recall that the vortex gas of the 2D classical XY model may be described by the sine-Gordon theory
\begin{align}
    {\cal L} = \frac{1}{4 \pi \beta} (\nabla \vartheta)^2 - \zeta ( e^{i \vartheta} + e^{-i \vartheta}).
\end{align}
This follows from the observation that the Boltzmann factor for a configuration of $m$ vortices and $n$ antivortices is computed by the path integral
\begin{multline}
    \int {\cal D} \vartheta \, e^{- \int {\cal L}_*} \prod\limits_{i=1}^{m} e^{i \vartheta(a_i)} \prod\limits_{j=1}^n e^{-i \vartheta(b_j)}
    \\= \delta_{m,n} \exp\biggl(-\beta \sum_{i,j} \log |a_i - b_j| + \beta \sum_{i<j}\log|a_i - a_j| +  \beta \sum_{i<j} |b_i - b_j|\biggr)
\end{multline}
where ${\cal L}_* = \frac{1}{4 \pi \beta} (\nabla \vartheta)^2$. 
Hence, the expansion in $\zeta$ of the partition function $Z = \int {\cal D} \vartheta \, e^{- \int {\cal L}}$ may be written as
\begin{multline}
    Z 
    = \sum_{m,n=0}^{\infty} \frac{\zeta^{m+n}}{m! n!} \int \prod_{i=1}^m d^2 a_i \prod_{j=1}^n d^2 b_j \,
    \\ \times \delta_{m,n} \exp\biggl( - \beta \sum_{i,j} \log|a_i - b_j| + \beta \sum_{i<j} \log |a_i - a_j| + \beta \sum_{i<j} \log |b_i - b_j| \biggr),
\end{multline}
showing that $Z$ is indeed the grand canonical partition function for the gas. $\beta$ is the inverse temperature and $\zeta$ is the fugacity. The operator $e^{i q \vartheta}$ (with $q \in \RR$) corresponds to a vortex of vorticity (or charge) $q$. 

We wish to do a similar thing here, that is, to write a field theoretic description for the gas of instanton and anti-instanton constituents in the $\CP^{N-1}$ model. 
In the remainder of this section, for clarity of the exposition, we shall give the development of this description only for the simplest case of $\CP^1$. 
We shall return to the general case of $\CP^{N-1}$ in the next section where a more technical aspect of the development, the cluster expansion, will be treated in more detail. 

In the $\CP^1$ model, we have only two kinds of vortices, $\vortex_1, \vortex_2$, and two kinds of antivortices, $\antivortex_1, \antivortex_2$. 
(For ease of language, and by analogy with the XY model reviewed above, let us agree from now on to refer to instanton constituents as `vortices' and anti-instanton constituents as `antivortices.')
However, since each of our vortices is supposed to carry a pair of decoupled charges, one real and one imaginary, we should make use of two scalar fields, $\varrho, \vartheta$. 
Then we can write a generalized vortex operator in the form
$e^{iq_{\text{im}} \varrho} e^{iq_{\text{re}}\vartheta}$
where $q_{\text{re}}, q_{\text{im}}$ are the real and imaginary charges of the vortex. 

Thus, we shall model the gas of these vortices and antivortices by a theory of a complex scalar field $y = \varrho - i \vartheta$, taking the Lagrangian preliminarily to be
\begin{align}
    {\cal L} &= K |\nabla y|^2 - \zeta(e^{-y} + e^{y}) - \zetabar ( e^{-\ybar} + e^{\ybar})
    \label{eq:SFT0}
\end{align}
Here, $K \equiv g^2 / 4\pi^2$ and $\zeta \equiv m_0 \mu e^{- (I - i \theta)/2}$, where $I = 2 \pi/g^2$ is the instanton action, $\theta$ is the topological angle, $\mu$ is the cutoff scale, and $m_0$ is a parameter with mass dimension one. 
We cannot really fix $m_0$ in our effective theory here, though we know that it must be nonzero and positive because the fugacity of our gas should be finite and positive when the topological angle is set to zero. Since the only dimensionful quantity in the theory is the strong scale $\Lambda$, we naturally expect $m_0= c_0 \Lambda$ where $c_0$ is some order one pure number.   
Later, we shall consider a $\CP^1$ model with fermions, namely, the deformation of the ${\cal N} = (2,2)$ supersymmetric $\CP^1$ model by a bare fermi mass term, and there we can show that $m_0$ is given by the fermion mass parameter.  

We shall assign to vortices and antivortices the following operators:
\begin{align}
    \vortex_1 \sim e^{-y}, 
    \quad \vortex_2 \sim e^y, 
    \quad \antivortex_1 \sim e^{-\ybar},
    \quad \antivortex_2 \sim e^{\ybar}.
    \label{eq:vortex-op-1}
\end{align}
To justify these assignments, we note that, with $\langle \ldots \rangle_*$ denoting the expectation value with respect to the Lagrangian ${\cal L}_* = K |\nabla y|^2$, we have
\begin{align}
    \langle \vortex_i(s) \vortex_j(t) \rangle_* = \langle \antivortex_i(s) \antivortex_j(t) \rangle_* = 0,
\end{align}
in accordance with the principle that vortices do not interact with other vortices, and likewise for antivortices. 
On the other hand, we have
\begin{align}
    \langle \vortex_i(a) \antivortex_j(b) \rangle_*
    = \langle e^{(-1)^i y(a)} e^{(-1)^j \ybar(b)} \rangle_* 
    = \exp\biggl( - (-1)^{i+j} \frac{\pi}{g^2} \log|a-b|^2 \biggr). 
    \label{eq:IntVVbar}
\end{align}
which implies that the vortex of type $i$ at position $a$ and an antivortex of type $j$ at position $b$ shall have an interaction equal to $(-1)^{i+j} \frac{\pi}{g^2} \log|a-b|^2$. 
This agrees perfectly with our general formula Eq.~\eqref{eq:constituent-interaction} for the interaction of a vortex--antivortex pair,\footnote{Note that for $N=2$, we have $\bm \nu_i \cdot \bm \nu_j = \delta_{ij} - \frac{1}{2} = \frac{1}{2} (-1)^{i+j}$.} and we have thus justified the operator assignments in \eqref{eq:vortex-op-1}.

It should be mentioned that the vortex operator assignments \eqref{eq:vortex-op-1} is not new. In the context of the ${\cal N} = (2,2)$ supersymmetric $\CP^1$ model, Ref.\cite{Hori:2000kt} argued that the supersymmetric completion of the vortex operator $e^{i \vartheta}$ requires a chiral twisted superfield $Y$, leading to the vortex operator $e^{-Y}$, where the lowest component of $Y$ is the cylinder-valued field $y = \varrho - i \vartheta$. But it should be stressed that the arguments of  Ref.\cite{Hori:2000kt} make no mention of the imaginary charges we have introduced in this work. Indeed, our approach, which up to this point has not relied on supersymmetry at all, boils down essentially to demanding perfect consistency with all the known classical properties of the instantons.

Applying a mean field analysis of \eqref{eq:SFT0}, we find classical solutions at  constant field configurations (mean field values):
\begin{align}
    e^{-y_{\rm cl}} = \pm 1.
\end{align}
The mean field solutions here imply a fractional theta angle dependence and will be related with the expected multibranched vacuum structure of the theory. 
The fugacities of $\vortex_1$ and $\vortex_2$ are proportional to $e^{-t/2}$. This means that the densities of $\vortex_1$ and $\vortex_2$ are equal at the mean field level. The combination $\mu e^{-{I (\mu)}/{2}} = \Lambda $ is the RG invariant strong scale.

While our story seems to be on the right track, there is an immediate problem with it. 
The potential generated by the proliferation of vortices, being the real part of an analytic function, is unbounded from below. 
In fact, the system is not even locally stable -- at the mean field values, there are always tachyonic fluctuations. 
This instability is related to the fact that there are imaginary charges in the system. 
But the theory will find ingenious ways to fix these problems itself, as we will see below. 

To see the local instability most clearly, set $\theta=0$ and expand the action around one of the mean field values at leading order in the fugacity. 
The Lagrangian written in terms of the real and imaginary parts of $y$ is 
\begin{align}
    {\cal L}&= K [(\nabla \varrho)^2  +  (\nabla \vartheta)^2 ] -  2 \zeta \cosh \varrho  \cos    \vartheta,  
    \label{eq:RIcoulomb1}  
\end{align}
where $\zeta = m_0 \Lambda$.
The two distinct mean field values according to the leading order Lagrangian \eqref{eq:RIcoulomb1} are located at  $(\varrho,  \vartheta)=(0,0), \; (0,\pi)$.  Expanding the potential to quadratic order, we find $\pm(\delta \varrho)^2 \mp (\delta \vartheta)^2$ around the two critical points. 
This means that one of $\varrho$ or $ \vartheta$ is always tachyonic. This tachyonic sickness 
cannot be cured at leading order in the cluster expansion. 

In the Coulomb gas language, it is easy to see physically why at leading order in fugacity the energy of this system should be unbounded from below. First of all, note that keeping only leading order terms in $\zeta$ corresponds to neglecting higher order effects in the cluster expansion, which have important qualitative  effects. $\vortex_i$ does not interact with $\vortex_j$ and  $\antivortex_i$ does not interact with  $\antivortex_j$. At leading order, $\vortex_1$s attract $\antivortex_1$s and an arbitrary number of them will clump. 
Likewise, $\vortex_2$s attract $\antivortex_2$s, and they will clump as well. Since $\vortex_1$s repel  $\antivortex_2$s and  $\vortex_2$s repel  $\antivortex_1$s, these two clumps will move away from each other within the volume of our subsystem. 
Since the energy of each clump is unbounded from below, so too is the energy of the generalized Coulomb gas (at first order in fugacity). 

A possible resolution appears in the second order in fugacity $\zeta$, where we have contributions from two-body clusters:
\begin{align}
    [\vortex_1 \antivortex_1],\ 
    [\vortex_2 \antivortex_2],\   
    [\vortex_1 \antivortex_2],\   
    [\vortex_2 \antivortex_1].
    \label{eq:bions}
\end{align}
These two-clusters (or bions) generate a potential which includes $ \cosh 2 \varrho ,  \cos  2  \vartheta$ terms. 
The inclusion of the operators corresponding to these two-clusters can in principle fix the instability problem that appears at first order in $\zeta$, as will be discussed below. 

\subsection{Help from mirror symmetry in the supersymmetric theory}

Despite the fact that the potential of \eqref{eq:SFT0} seems to imply an instability at leading order in $\zeta$, we can actually confirm by other means that this nonperturbatively generated potential is actually correct, albeit incomplete and part of a larger story. 
Here, we shall use some arguments from mirror symmetry of the supersymmetric version of the $\CP^1$ model to reproduce \eqref{eq:SFT0}, among other things.  
According to mirror symmetry, the supersymmetric ${\cal N}=(2,2)$  $\CP^1$ model is dual to a theory of a complex (twisted) chiral superfield $Y$ (or ``Landau--Ginzburg model'') with superpotential \cite{Hori:2000kt}
\begin{align}
    \widetilde{W}\,=2 \Lambda  \cosh Y =  \Lambda ( e^{-Y}  +   e^{+Y}) .
\end{align}
The terms that arise from the superpotential  and $F$-term are
\begin{multline}  
    \tfrac{1}{2} \int d^2 \widetilde{\theta}  \; \widetilde{W} +  \cc  + FF^* 
    \\= -  \Lambda (   e^{- y} \overline  \chi_{+} \chi_{-} +  e^{+ y} 
    \overline  \chi_{+} \chi_{-}  +  e^{- \ybar} \overline  \chi_{-} \chi_{+} + e^{+ \ybar}   
    \overline  \chi_{-} \chi_{+} ) 
    + 2  \Lambda^2 (  \cosh 2\varrho  -   \cos  2\vartheta),
    \label{eq:susy}
\end{multline}
where we have eliminated the non-dynamical field $F$ using the Euler--Lagrange equation.     
The term proportional to $\Lambda$ in \eqref{eq:susy} just arises from the same set of vortices as in \eqref{eq:vortex-op-1}, but with the attachments of the fermi zero modes: 
\begin{align}
    \vortex_1  &\sim  e^{- y} \overline  \chi_{+} \chi_{-}, \qquad   \vortex_2  \sim  e^{+ y}
    \overline  \chi_{+} \chi_{-}, \cr
    \antivortex_1  &\sim  e^{- \ybar} \overline  \chi_{-} \chi_{+}, \qquad   \antivortex_2  \sim  e^{+ \ybar} 
    \overline  \chi_{-} \chi_{+}.
    \label{eq:Vortexsusy}
\end{align} 
It must be admitted that in our perspective, it is difficult to rationalize these attachments of fermi bilinears to the vortex operators. 
We have emphasized that the instanton constituents are merely fictitious, and thus it does not quite make sense to say that they support fermi zero modes.
A rationale within our framework, while not completely satisfactory, can be given based on the following considerations. 
It is known that the 2D instanton has four zero modes: ${\cal I} \sim (\overline  \chi_{+} \chi_{-})^2$. 
The fermi fields attached to the vortex operators $\vortex_1, \vortex_2$ should be such that ${\cal I} = \vortex_1 \vortex_2$.
Given this, we can fix the fermi bilinears attached to $\vortex_1, \vortex_2$ by fermi number and Lorentz invariance, and by demanding that the transformation $y \to -y, \chi_{\pm} \to - \chi_{\pm}, \chibar_{\pm} \to - \chibar_{\pm}$ effects the interchange $\vortex_1 \leftrightarrow \vortex_2$ of the vortex operators.

The instanton, as implied by the ABJ anomaly, explicitly breaks the classical $\U(1)_A$ chiral symmetry down to  $\ZZ_4$. 
Since $\ZZ_4$ is an exact symmetry of the microscopic theory, it must also be an exact symmetry of the dual formulation. 
To see this, recall that to all orders in perturbation theory, $\vartheta$ acquires a continuous shift symmetry $\U(1)_J$. 
This is explicitly broken by the vortex operators. However, a certain subgroup $\ZZ_4$ of $\U(1)_J \times \ZZ_4$ survives. 
\begin{align}
    \ZZ_4\colon \quad
    \overline  \chi_{+} \chi_{-} \to - \overline  \chi_{+} \chi_{-},\quad
    e^{- \varrho + i\vartheta}  \to - e^{- \varrho + i\vartheta}
    \label{discretesym}
\end{align}
Hence, $\vortex_i, \antivortex_i$ are invariant under the $\ZZ_4$ symmetry. 

Because of the nonrenormalization of the superpotential, some of the physics can be studied exactly in the mean field approximation. 
Since $\vartheta$ is an angle-valued scalar, the term $- \cos  2 \vartheta$ leads to two minima: 
\begin{align}
    \langle  e^{-Y} \rangle = \langle  e^{-\varrho + i\vartheta} \rangle = \pm 1   
    \label{eq:condensate}
\end{align}
Note that the dynamical chiral symmetry breaking in the original $\CP^{1}$ model maps 
into spontaneous breaking by a non-perturbatively generated 
potential visible at \emph{tree-level} in the dual formulation. 
This is often an advantage of duality: 
A phenomenon that is not very transparent in the original formulation is often manifest at the tree-level in the dual formulation.
Expanding around either vacuum, we see that all components of $Y$ acquire  
the same mass, as implied by ${\cal N}=(2,2)$ supersymmetry. 
Note that the mass of the scalar excitations is due to two-cluster or bion effects \eqref{eq:bions} which generate a bosonic potential, $2 \Lambda^2 (  \cosh 2 \varrho  -   \cos  2   \vartheta)$. 
The mass for the fermions is due to the formation of the nonperturbative condensate \eqref{eq:condensate}.  
The  vortex induced operators once the chiral symmetry is broken by the condensate, maps to fermi mass terms plus interactions
\begin{align}
    \Lambda   \langle   e^{- \varrho + i\vartheta} \rangle  \;  {\overline  \chi_{+}} \chi_{-} + \ldots = \pm  \Lambda  \; {\overline    \chi_{+}} \chi_{-} + \ldots   
\end{align}
This is the generation of the fermion mass by chiral symmetry breaking in the dual formulation of the theory. 

If we turn on  a soft  chiral symmetry breaking  mass term for the fermions,
\begin{align}
    \Delta {\cal L} = m \chibar_- \chi_+ + \cc, 
\end{align}
the fermi zero modes attached to the vortices get lifted, and the vortex operators become purely bosonic. In consequence, the bosonic potential is modified (assuming for the moment that $m$ is a real parameter) to
\begin{align}  
V(\varrho, \vartheta) &= - 2 m  \Lambda   \cosh \varrho  \cos    \vartheta     + 2\Lambda^2 (  \cosh 2 \varrho  -   \cos  2   \vartheta).   \cr 
&\leftrightarrow  -  \left( \vortex_1 + \vortex_2 +   \antivortex_1 +  \antivortex_2 +    [ \antivortex_1   \vortex_1]  +
[ \antivortex_2   \vortex_2] +  [ \antivortex_1   \vortex_2] + [ \antivortex_2   \vortex_1] \right)
\end{align}
The first term in the potential   $ - 2 m  \Lambda   \cosh \varrho  \cos    \vartheta $ matches the result we obtained via our refined instanton analysis, given in \eqref{eq:SFT0}, as contributions of vortices and antivortices.
The second term can be interpreted as contribution of two-clusters $ [ \vortex_i  \overline   \vortex_j]  $.  
In particular, it is important to note the crucial sign difference between the fugacities.  
The fugacity of $ [\vortex_1   \antivortex_1]  $ is opposite in sign to that of 
$[ \vortex_1   \antivortex_2]  $, i.e, 
\begin{align}
\zeta_{[  \vortex_1   \overline  \vortex_1]} = \zeta_{[ \vortex_2  \overline  \vortex_2]} =   e^{i\pi}  \zeta_{[  \vortex_1  \antivortex_2]} 
= e^{i\pi}  \zeta_{[  \vortex_2  \antivortex_1]} 
\end{align} 
The modulus  of the fugacity, which correspond to the density, is the same for both configurations. 
If the fugacities of ${[  \vortex_1   \antivortex_1]}$ and  ${[  \vortex_2   \antivortex_2]}$ were both positive, then we would have instead the term $- \cosh 2 \varrho$ in potential and the theory would suffer an instability.  
It is the relative $\pi$ phase in the fugacities that comes to the rescue.\footnote{This extra angle is called hidden topological angle \cite{Behtash:2015kna}. 
It has similar manifestations in gauge theories on $\RR^3 \times \SS^1$ as well as in quantum mechanical systems \cite{Behtash:2015loa} where it is calculated explicitly by using Lefshetz thimbles associated with steepest descent cycles.} 

Thus, we see that in the soft mass deformation of the supersymmetric $\CP^1$ model, the seemingly fatal instability is cured by the two-cluster terms. In the next section, we shall give a detailed computation of these contributions.

\section{Subextensive contributions in the cluster expansion}

At leading order in fugacity, the refined instanton gas seems to be plagued by a fatal instability. 
We have argued that the instability could be fixed by taking into account the subextensive contributions, or two-body clusters, in the cluster expansion. 
We saw that this is precisely what happens in the supersymmetric theory according to mirror symmetry. 
In this section, we give a direct calculation of these two-cluster effects in the supersymmetric theory. 
As a byproduct, we obtain a new derivation of the exact superpotential \cite{Hori:2000kt} using nonsupersymmetric tools. 

\subsection{Vortex operators for the \texorpdfstring{$\CP^{N-1}$}{CPN-1} model}

We shall work in the general case of $\CP^{N-1}$, so to this case we must generalize the discussion of the previous section. 
For the bosonic model, we describe the generalized Coulomb gas using $N-1$ complex scalar fields $\bm y = (y_1,\ldots,y_{N-1})$ and taking as a preliminary Lagrangian
\begin{align}
    {\cal L} = K|\nabla \bm y|^2 - \zeta \sum_{i=1}^N e^{-\bm \nu_i \cdot \bm y} - \zetabar \sum_{i=1}^N e^{-\bm \nu_i \cdot \bm \ybar}
\end{align}
where $K = Ng^2/16 \pi^2$ and $\zeta = m_0 \Lambda e^{i \theta/N}$. The various vortices and antivortices $\vortex_i, \antivortex_i$ ($i=1,\ldots,N$) are assigned the operators
\begin{align}
    \vortex_i \sim e^{- \bm \nu_i \cdot \bm y}, \quad
    \antivortex_i \sim e^{- \bm \nu_i \cdot \bm \ybar}.
    \label{eq:vortex-op-bose-general}
\end{align}
One can easily check, as we had done previously in the case of $\CP^1$, that this operator assignment is perfectly consistent with the requirement that vortices do not interact with vortices, and that the interaction between the $i$-th vortex and the $j$-th antivortex be given by \eqref{eq:constituent-interaction}. 

In the supersymmetric theory, we assume that the vortex operators should be accompanied by fermi bilinears:
\begin{align}
    \vortex_i \sim e^{-\bm \nu_i \cdot \bm y}(\bm \nu_i \cdot \bm \chibar_+) (\bm \nu_i \cdot \bm \chi_-), 
    \quad \antivortex_i \sim e^{- \bm \nu_i \cdot \bm \ybar}(\bm \nu_i \cdot \bm \chibar_-) (\bm \nu_i \cdot \bm \chi_+).
    \label{eq:vortex-op-susy-general}
\end{align}
As in the supersymmetric $\CP^1$ case discussed previously, we cannot fully justify this step within our framework. Our rationale is based on the fact that the full instanton or anti-instanton should have $2N$ fermi zero modes, and thus the instanton operator, which arises from taking the product $\prod_{i=1}^N \vortex_i$, should come with $2N$ fermi fields, schematically, ${\cal I} \sim \prod_{i=1}^N (\bm \nu_i \cdot \bm \chibar_+) (\bm \nu_i \cdot \bm \chi_-)$. Given this, the form of the fermi bilinears attached to the various vortex operators is dictated by fermi number conservation, Lorentz invariance, and demanding that any transformation of the supermultiplet $\bm Y$ according to the standard representation of the symmetric group $S_N$ effects a permutation of the vortex operators: 
\begin{align}
    S_N \ni \sigma \colon \bm Y \mapsto D(\sigma) \cdot \bm Y \implies \vortex_i \mapsto \vortex_{\sigma(i)},
\end{align}
where $D$ is the standard $(N-1)$-dimensional representation of $S_N$. 
In any case, the mirror symmetry result established in \cite{Hori:2000kt} justifies our assumption a posteriori. 
The dual Landau--Ginsburg model is given by a $N-1$ complex (twisted) chiral superfields $\bm Y$ with superpotential
\begin{align}
    \widetilde{W} = \Lambda \sum_{i=1}^N e^{- \bm \nu_i \cdot \bm Y},
\end{align}
and this leads to the appearance of the vortex operators \eqref{eq:vortex-op-susy-general}. 
From this, we can immediately deduce the bosonic potential:
\begin{align}
{\textstyle \frac{2}{\Lambda^2} } {V} (y_i) \,=  {\textstyle \frac{1}{\Lambda^2} }  \sum_{i=1}^{N}   \left| \frac{ \partial W}{\partial y_i} \right|^2  &=    \sum_{i=1}^{N}  \left|    e^{-  {\bm \nu}_i \bm y}  
-\frac{1}{N}  \sum_{ j=1 }^{N}   e^{-  {\bm \nu}_j \bm y } \right|^2   \cr
&= \sum_{i=1}^{N}     e^{-  {\bm \nu}_i (\bm y + \bm \ybar)}  
-\frac{1}{N}  \sum_{i, j }   e^{-  {\bm \nu}_i \bm y -  {\bm \nu}_j  \bm \ybar}    \cr 
&= \left( \frac{N-1}{N} \right)  \sum_{i=1}^{N}    e^{-  {\bm \nu}_i (\bm y + \bm \ybar)}  
-   \frac{1}{N}  \sum_{i \neq  j }   e^{-  {\bm \nu}_i \bm y -  {\bm \nu}_j  \bm \ybar}  \cr
&=  \sum_{i=1}^{N}  {\cal B}_{ii} + \sum_{i \neq  j }   {\cal B}_{ij}
\label{bospot2}
\end{align} 
As we will now show, this bosonic potential obtained from the superpotential is precisely reproduced via cluster expansion and quasizero mode Lefschetz thimble integration. 

\subsection{Bosonic potential from cluster expansion and Lefshetz thimble integration}

\label{sec:bions}

The path integral of the effective theory to first order in the cluster expansion in terms of  constituents 
is given by 
\begin{align}
Z= \int {\cal D} [\text{fields}] \, e^{ - \int d^2x ( {\cal L}_0 +  {\cal L}_1) }
\end{align}
where $Z$ is  the partition function for a the grand canonical ensemble. 
From this, we can actually deduce terms at second order in cluster expansion. 
By expanding the exponential $e^{ - \int d^2x {\cal L}_1} = 1   - \int d^2x {\cal L}_1 + \frac{1}{2} 
 (\int d^2x   {\cal L}_1)^2$ 
we find terms at second order such as  
\begin{align}
    \sum_{ij} \int d^2a \, d^2b \,  \vortex_{i} (a)  \antivortex_{j} (b) 
    =  \sum_{ij} \int d^2R \, d^2r \, \vortex_{i} (R + \tfrac{1}{2}r)  \antivortex_{j} (R - \tfrac{1}{2}r)
\end{align}
where we have pulled out the integration over the exact ``center''  zero mode $R$.
This induces an effective second-order term in the cluster expansion: 
\begin{align}
{\cal L}_2  \supset  -  \sum_{ij} {\cal B}_{ij} = -  \sum_{ij} \int  d^2r\,  \Bigl\langle \vortex_{i} (R + \tfrac{1}{2}r)  \antivortex_{j} (R - \tfrac{1}{2}r)  \Bigr\rangle
\label{QZMint}
\end{align}
where the $\langle\ldots\rangle$ stands for a connected correlator in the perturbative vacuum. 
If the correlator is mainly supported at separations $r < r_b$ for some length scale $r_b$, then integrating out those length scales, we can view the  ${\cal B}_{ij}$ as independent local operators in an effective action. 
We refer to these two-cluster operators as bion operators.

Note that since $\langle \vortex_{i} (R + \tfrac{1}{2}r)   \vortex_{j} (R - \tfrac{1}{2}r)  \rangle =0$ in the perturbative vacuum -- reflecting the fact there is no classical interaction between $\vortex_{i}$  and $\vortex_{j}$ -- there is no clustering between vortex--vortex (or antivortex--antivortex) pairs at the classical level. 
The two-cluster terms are generated only by vortex--antivortex pairs.

To find the correlated amplitude $  [ \vortex_i \antivortex_j ] $, we need to  perform the integration over the relative coordinate $r$. 
This is equivalent to integration over the quasi-zero mode direction. 
However, an unusual property of the Coulomb potential in two dimensions is that it blows up at infinity. 
Therefore, we need to be more careful with it. 
In fact, we will do computations below in arbitrary $d$ dimensions, and take $d \to 2$ limit at the end of the computation.

The correlator of $\vortex_i(r)$ and $\antivortex_j(0) $ which helps us to identify both classical and  fermion zero mode induced (quantum) interactions between two vortices can be written as
\begin{align}
    \Big \langle \vortex_i(r) \antivortex_j(0)  \Big  \rangle   
    &= \Big  \langle  e^{- {\bm \nu_i} \cdot \bm y }     ({\bm \nu_i} \cdot \bm \chibar_{+})     ({\bm \nu_i} \cdot \bm \chi_{-}) (r)   \, \,  e^{- {\bm \nu_j} \cdot \bm \ybar  }     ({\bm \nu_j} \cdot \bm \chi_{+})     ({\bm \nu_j}  \cdot \bm \chibar_{-})   (0) \Big \rangle    \nonumber\\
    &=  e^{- V_{ij}(r) } \frac{({\bm \nu}_i \cdot {\bm \nu}_j )^2 }{|r|^2}
    \label{int-susy}
\end{align} 
Here, $V_{ij}(r)$ is the classical interaction between two vortices given in  
\eqref{eq:constituent-interaction}, and $1/r = \langle \chi_+(r) \chibar_+(0) \rangle$ is the massless fermion propagator in 2D, which we interpret as a contribution $\log r$ to the interaction due to fermi zero mode exchange. 
Note that unlike classical interactions, which are $O(1/g^2)$, the fermi zero mode exchange-induced interaction does not have the coupling in front. 
It is $O(g^0)$ and needs to be interpreted as a 1-loop level interaction. 
These are familiar both from gauge theories on $\RR^3 \times \SS^1$ and supersymmetric  quantum mechanical systems.

In supersymmetric quantum field theories and supersymmetric quantum mechanics, the interaction between instanton--anti-instanton pairs can written in a particularly simple way:
\begin{align}
\Big \langle \vortex_i(r) \antivortex_j(0)  \Big  \rangle   = \nabla^2 
  e^{- U_{ij}(r) } 
   \label{int-susy2}
\end{align} 
A formal way to see this generally is to elevate the separation quasizero mode coordinate to a supersymmetry invariant coordinate by introducing accompanying Grassmann coordinates and then integrating over them \cite{Balitsky:1985in}. 
But it is also quite easy to check that \eqref{int-susy2} is equal to \eqref{int-susy}.

Therefore, the bion operator may be written as 
\begin{align}
{\cal B}_{ij} = [\vortex_i]  [\antivortex_j  ] \times I_{ij} (\lambda) 
\label{bionop}
\end{align}
where  $I_{ij} (\lambda)$ is the quasi-zero mode integral. Using Eqs. \eqref{int-susy2} and \eqref{QZMint}, we can express the integral for a $d$-dimensional supersymmetric completion of the Coulomb gas as 
\begin{align} 
    I_{ij} (\lambda)  &= \int d^d r   \;   \nabla^2 e^{-  V_{ij}(r) },  \\
    V_{ij}(r) &\equiv  \frac{1}{\lambda_{ij}} \frac{ \Gamma ( \frac{d-2}{2})}{4 \pi^{d/2}} \frac{1}{r^{d-2}}, 
\end{align} 
where $\lambda_{ij}$ is our substitute for coupling:
\begin{align}
 \lambda_{ij}
=  \left\{
 \begin{array} {ll} 
   -\frac {NK} {N-1}   &  \qquad  {\rm attractive \; for \;}  \;  i=j   \cr  \cr
 NK        &  \qquad  {\rm repulsive  \; for \;}  \;  i\neq j  
 \end{array}   
 \right.
 \label{class-int2}
 \end{align}

This integral can be reduced to a one-dimensional integral of a total derivative by using simple steps: 
\begin{align}
    I_{ij} (\lambda) 
    &= \Omega_d   \int_0^{\infty} d r \, r^{d-1}  \frac{1}{r^{d-1}}\frac{\partial}{  \partial r}  \left( r^{d-1}   \frac{\partial}{  \partial r}    e^{- \frac{1}{\lambda}  V(r) } \right) \nonumber\\
    &= \frac{\Omega_d}{\lambda}    \frac{ \Gamma \left( \frac{d+2}{2} \right)  }{2 \pi^{\frac{d}{2}}}    
    \int_{0}^{\infty} dr  \, \frac{\partial}{  \partial r}    e^{- \frac{1}{\lambda}  V(r) } 
\end{align}
Although the integrand is a total derivative so that the integral is determined by the end points, some care is still required. 
The integral over $r$ is naively over the positive real axis, $r \in [0,+\infty)$.
However, these quasi--zero mode integrals need to be done by finding the critical points of the classical interaction, and then identifying the associated steepest  descent cycles (or Lefschetz thimbles) ${\cal J}(\lambda)$  in the complex $r$ plane. 
In the present case, as in the case of a 3d Coulomb gas, the critical point is at infinity. 
The cycle starts there and ends at zero, which is a pole for $d=3, 4$, but a branch point for generic $d$. 
In general, the Lefschetz thimble ${\cal J}(\lambda)$ changes as  $\arg \lambda$ is varied. 
For  $\arg \lambda =0$, the cycle ${\cal J}(\lambda)$ enters zero from the direction $\arg r = 0$, and the Lefschetz thimble is in fact the naive integration contour, the positive real axis. 
On the other hand, for $\arg \lambda=\pi$, the cycle  ${\cal J}(\lambda)$ enters zero from the direction $\arg r = \pi$, and in fact the integration contour is the \emph{negative} real axis.
The upshot is that in both cases $\arg \lambda = 0, \pi$, we have $e^{-\frac{1}{\lambda} V(r)} \to 0$ as $r \to 0$ along the appropriate thimble ${\cal J}(\lambda)$, so that the contribution of the end point at $r=0$ is zero. 
Thus, the quasi--zero mode integral is determined by the end point at $r = \infty$,
\begin{align}
    I_{ij} &= \frac{\Omega_d}{\lambda}    \frac{ \Gamma \left( \frac{d+2}{2} \right)  }{2 \pi^{\frac{d}{2}}}    
    \int_{{\cal J}(\lambda)} dr  \;   \frac{\partial}{  \partial r}    e^{- \frac{1}{\lambda}  V(r) }   \nonumber\\
    &=   \frac{  \Omega_d   \Gamma \left( \frac{d+2}{2} \right)  }{2 \pi^{\frac{d}{2}}} \times     \frac{1}{\lambda_{ij}},
\end{align}
and we reach the elegant conclusion that the quasizero mode integral in the repulsive and attactive cases are related by a simple formula:
\begin{align}
    {I}_{ii} =   (N-1) e^{i \pi}   I_{i j} \qquad (i\neq j).
\end{align}

We associate the bosonic potential with two-cluster operators (or bions)  in the cluster expansion. 
These operators can be written as
\begin{align}
 {\cal B}_{ii} &= [ \vortex_i \antivortex_i ]   =  \left(   \frac{N-1}{N} \right)  e^{-2 \frac{4 \pi }{g^2N} }  e^{i \pi} e^{-  {\bm \nu}_i (y + y^*)}   \cr
{\cal B}_{ij}  &=   [ \vortex_i \antivortex_j ] = \frac{1}{N}    e^{-2 \frac{4 \pi }{g^2N}  } 
    e^{-  {\bm \nu}_i y -  {\bm \nu}_j  y^*} \qquad (i \neq j)   
    \label{bions2}
\end{align}
Note that the contribution of ${\cal B}_{ii}$ to vacuum energy density is negative,   while the contribution of the ${\cal B}_{ij}$ is positive.  
In particular, the relation between vacuum values of the bion amplitudes is 
\begin{align}
{\cal B}_{ii}^{\rm vac}  =   (N-1) e^{i \pi}   {\cal B}_{ij}^{\rm vac} 
\end{align}
which differ both in sign and magnitude. 
In a grand canonical ensemble, the fugacity associated with ${\cal B}_{ij}$ is positive, while the fugacity associated with ${\cal B}_{ii}$ is negative. 
In the vacuum, $y=0$,  the two contributions balance out to cancel exactly.
 \begin{align}
 \frac{1}{\Lambda^2} {\cal E}_{\rm vac} =   \left( \frac{N-1}{N} \right)     \sum_{i=1}^{N}  1
-   \frac{1}{N}  \sum_{i \neq  j }    1 =  \left( \frac{N-1}{N} \right) N  -    \frac{1}{N}  N(N-1) =0
\label{vacen} 
\end{align}
giving a zero non-perturbative vacuum energy density in the supersymmetric theory.

 \subsection{What objects enter into the statistical field theory?}
Both for the bosonic and supersymmetric theories, what enters into the effective theory (in the dual formulation) up to the second order in the cluster expansion are the operators associated with generalized vortices and two-clusters thereof. 
The charges of these objects are  valued in the $\SU(N)$ weight lattice and its imaginary copy,  $i   \weightlattice \oplus \weightlattice$, and are given by:
\begin{align} 
    \left.
    \begin{array} {ll}
        { \cal V}_i\colon  &   \qquad   (i  \nu_i,  \nu_i)   \cr
        \overline { \cal V}_i\colon  &   \qquad    (i  \nu_i,  - \nu_i)  \cr
        {\cal B}_{ii} = [ {\cal V}_i \overline {\cal V}_i ]  \colon  & \qquad  (i  2  \nu_i,  0) \cr
        {\cal B}_{ij}  =   [ {\cal V}_i \overline {\cal V}_j ] \colon &  \qquad  (i  (\nu_i + \nu_j),  (\nu_i - \nu_j)  ) \cr 
    \end{array} \right\} \in  i   \weightlattice \oplus \weightlattice 
    \label{list-cluster}
\end{align}
The supersymmetric model written in the component notation takes the form:
\begin{align}   
    {\cal L} 
    &= {\frac{N g^2}{16 \pi^2} }   [|\nabla \bm y |^2  + i  \bm \chibar_{-} (\partial_0 + \partial_1)  \bm \chi_{-}  +  i  \bm \chibar_{+} (\partial_0 - \partial_1)  \bm \chi_{+}    ]   \cr 
    &\quad\,-  \Lambda \sum_{i=1}^{N}   e^{- {\bm \nu_i} \bm y }     ({\bm \nu_i} \bm \chibar_{+})     ({\bm \nu_i}  \bm \chi_{-})   + {\rm {c.c.}} \cr 
    &\quad\,     +  \frac{\Lambda^2}{2} \left( \frac{N-1}{N} \right)   \sum_{i=1}^{N}    e^{-  {\bm \nu}_i (\bm y + \bm \ybar)}  
-    \frac{\Lambda^2}{2}   \frac{1 }{N}  \sum_{i \neq  j }   e^{-  {\bm \nu}_i \bm y -  {\bm \nu}_j  \bm \ybar}  
\label{componentdual}
\end{align}
and physically, it corresponds to the proliferation of the charges listed in \eqref{list-cluster}.\footnote{We have assumed here that the Kahler metric is trivial. This was necessary as a starting part for our analysis. We expect of course that quantum effects will renormalize the Kahler potential in a complicated way.}

The dynamical  breaking of the discrete chiral symmetry $\ZZ_{2N}$ in the original $\CP^{N-1}$ maps into spontaneous breaking by a non-perturbatively generated 
\emph{tree level} potential in the dual formulation.  
The bosonic potential  in \eqref{componentdual} has $N$ isolated  minima,  
\begin{align}
    \langle k |  e^{-  {\bm \nu}_i \bm y} |k \rangle  = e^{i  \frac{2 \pi k}{N}} \qquad (k=0, \ldots, N-1),
\label{csbvacua}
\end{align}
which corresponds to the spontaneous breaking of chiral symmetry $\ZZ_{2N} \rightarrow \ZZ_{2}$. 
Furthermore, the fact these minima are isolated implies the system is gapped. 
The discrete chiral symmetry breaking must generate a mass for the fermions. 
Indeed, using the vacuum value of $\bm y$ in the middle line of \eqref{componentdual}, we obtain 
\begin{align}  
    \Lambda \sum_{i=1}^{N} ({\bm \nu_i} . \bm \chibar_{+}) ({\bm \nu_i}.  \bm \chi_{-})     =    \Lambda  \sum_{r,s=1}^{N-1} \chibar_{+r}  M_{rs} \chi_{-s}
\end{align}
where the matrix $M_{rs}$is given by 
\begin{align}
    M = \sum_{i=1}^N \bm \nu_i \, \bm \nu_i^{\rm T}.
\end{align}
Because $M$ is invariant under the action of the permutation group in the sense that $D(\sigma)^{\rm T} M D(\sigma)$ ($D$ is the $N-1$ dimensional standard representation of $S_N$ and $\sigma \in S_N$), it must be proportional to the identity matrix. It follows that all fermions acquire the same mass. 



Turning on a soft supersymmetry breaking mass term perturbation
\begin{align}
    \Delta {\cal L}_m=  m \sum_{i=1}^N (\bm \nu_i.\bm \chibar_{+} ) (\bm \nu_i . \bm \chi_{-}) + \cc,
\end{align}
one can lift  the zero modes of the vortex operators. In consequence, this perturbation contributes to the vacuum energy density:
\begin{align} 
{\cal E}(\theta)=  \min_{k}  {\cal E}_k(\theta) = \min_{k}    -2  m \Lambda N   \cos \frac{\theta + 2 \pi k }{N} 
\label{branches}
\end{align}
This generates multi-branched  structure for the vacuum energy density.   The vacuum is unique for $\theta \notin  (2n+1) \ZZ $, and is given by the branch label $k^*$ 
over which  ${\cal E}_k(\theta)$ is the minimum for a given value of  $\theta$.
For  $-\pi < \theta < \pi$, the vacuum branch label is $k=0$.  
For  $\pi < \theta < 3 \pi$, the vacuum branch label is $k=-1$.
This implies a non-analytic behavior and a  phase transition at $\theta=\pi$.  
At $\theta=\pi$,  the vacuum is doubly degenerate. 
 
The order parameter for CP symmetry is the topological charge density $f/2\pi$. 
Its vacuum expectation value is given by 
\begin{align} 
    \biggl\langle  \frac{f}{2 \pi}  \biggr\rangle =   m \Lambda N  \sin  \frac{\theta + 2 \pi k_{*} }{N} 
\end{align}
where  $k_{*} (\theta)$  is the vacuum branch for a given value of theta, e,g.    $k_{*} (\theta) = 0 $ for $|\theta| < \pi$, and 
$k_{*} (\theta) = -1 $  for  $\pi < \theta < 3 \pi$, implying that at 
$\theta=\pi$, $\langle  f/2\pi \rangle =    \pm m \Lambda  \sin  \frac{  \pi }{N} $, and spontaneous breaking of the CP symmetry.

\section{Soliton spectrum of the dual theory}

The superpotential of the LG theory has allowed us to deduce the existence of a mass gap, though we are not able to compute the masses of the elementary quanta in the dual description, due to our lack of control over the Kahler potential.
On the other hand, the soliton spectrum in the LG formulation, which corresponds to the elementary quanta in the original formulation, is completely determined and is in fact well-known \cite{Cecotti:1992qh, Cecotti:1992rm}. 
Below, after briefly reviewing it, we move to the more interesting question of the spectrum of the nonsupersymmetric theory. 
This will give us valuable information about the states of the original $\CP^{N-1}$ model, and demonstrate the lack of confinement in the supersymmetric model and the presence thereof in the nonsupersymmetric one.\footnote{In this paper, we use ``confinement'' in a looser sense than is standard these days. The standard definition of confinement relies on the existence of a one-form symmetry, which is not present in the sigma models studied here. By ``confinement'' here, we mean the absence of asymptotic states carrying gauge charge.} 

In the supersymmetric model, the BPS solitons are static solutions of the equations of motions 
interpolating between different vacua, $\bm y = \bm y_0$ and $\bm y= \bm y_k$: 
\begin{align}
    \frac{d y^r}{d x}  =  \frac{\alpha}{2} g^{r \sbar} \frac{\del \Wbar}{\del \ybar^{\sbar}}, \quad \text{where} \quad \alpha \equiv \frac{W(y_k) - W(y_0)}{|W(y_k) - W(y_0)|}. 
    \label{BPS}
\end{align}
In \eqref{csbvacua}, we denoted the chiral symmetry breaking vacua by $|k \rangle$.  
The total number of solitons interpolating between $|0\rangle$ and $|k \rangle$ is equal to the number of independent solutions of \eqref{BPS}. 
For example, between adjacent vacua  $|k=0 \rangle$ and $|k=1 \rangle$ there are $N$ possible paths:
\begin{align}
    y(-\infty)=0,  \qquad y(\infty)= {\nu_1, \ldots,  \nu_N}  
\end{align}
where $\nu_i\  (i=1, \ldots, N)$ are the weights of the defining representation of $\SU(N)$.  
Indeed, the supersymmetric theory does not confine.  
The elementary quanta  $\phi = (\phi_1, \ldots,    \phi_N)$ in the original formulation \eqref{eq:gauged1} becomes gapped and uncontrained as shown by Witten \cite{Witten:1978bc}. 
The soliton in the dual formulation is the deconfined gapped quanta in the original formulation of the theory. 

The solitons connecting vacua that are $k$ units apart, $|q\rangle$ to $|q+k \rangle$, have masses and degeneracies given by 
\begin{align}
M_{k}&= | W( y(\infty)) - W( y(-\infty))| = N \Lambda |e^{i \frac{2 \pi k }{N}} -1 | = \Lambda 2N \sin \frac{\pi k }{N},  \cr  \qquad {\rm deg}_k &= \frac{N!}{k!(N-k)!}.
\end{align}
These states generate the $k$-index antisymmetric representation of $\SU(N)$,  
$\phi_{ \lbrack i_1 } \cdots \phi_{i_k\rbrack}$, with  dimension  $\frac{N!}{k!(N-k)!}$. 
The mass of the minimal solitons $M_{1} = \Lambda 2N \sin \frac{\pi  }{N} $ in the dual description corresponds to the mass of the elementary quanta in the original description.

\begin{figure}[t]
\vspace{-1.5cm}
\begin{center}
\includegraphics[width = 0.9\textwidth]{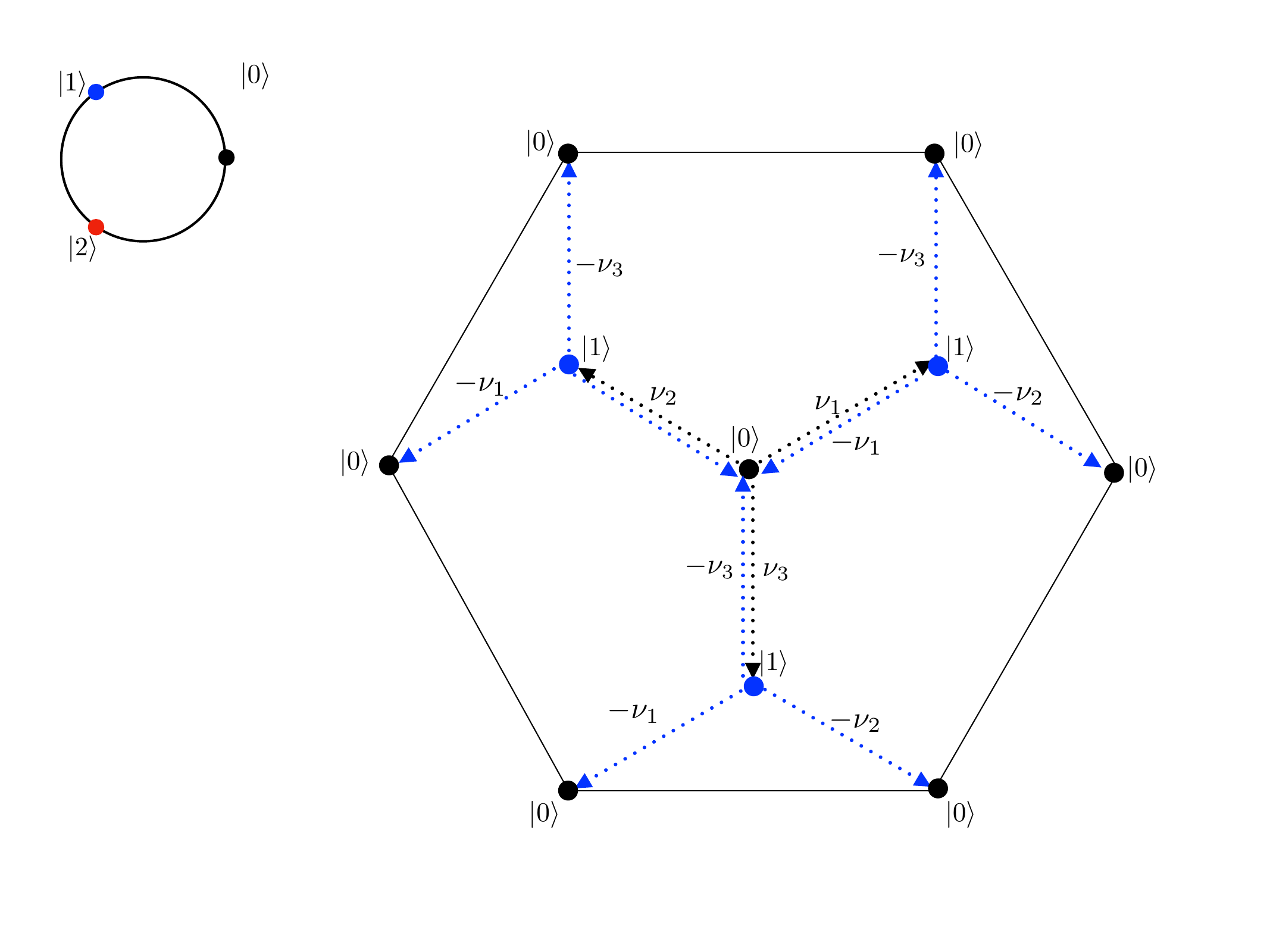}
\vspace{-1.5cm}
\caption{The minimal BPS solitons in the supersymmetric theory interpolate between adjacent vacua and furnish the defining representation of $\SU(N)$ (here, $N=3$). 
When supersymmetry is softly broken, the vacuum degeneracy is lifted, and a multi-branched vacuum structure obtains. 
In this case, these would-be domain wall costs infinite energy, but soliton--antisoliton pairs become stable and form the direct sum of the adjoint and singlet representations.
The soliton--antisoliton pair is held together by a linearly confining potential. 
 }
\label{fig:SAS}
\end{center}
\end{figure}

\begin{figure}[t]
\begin{center}
 \hspace*{3.0cm} 
 \includegraphics[width = 0.9\textwidth]{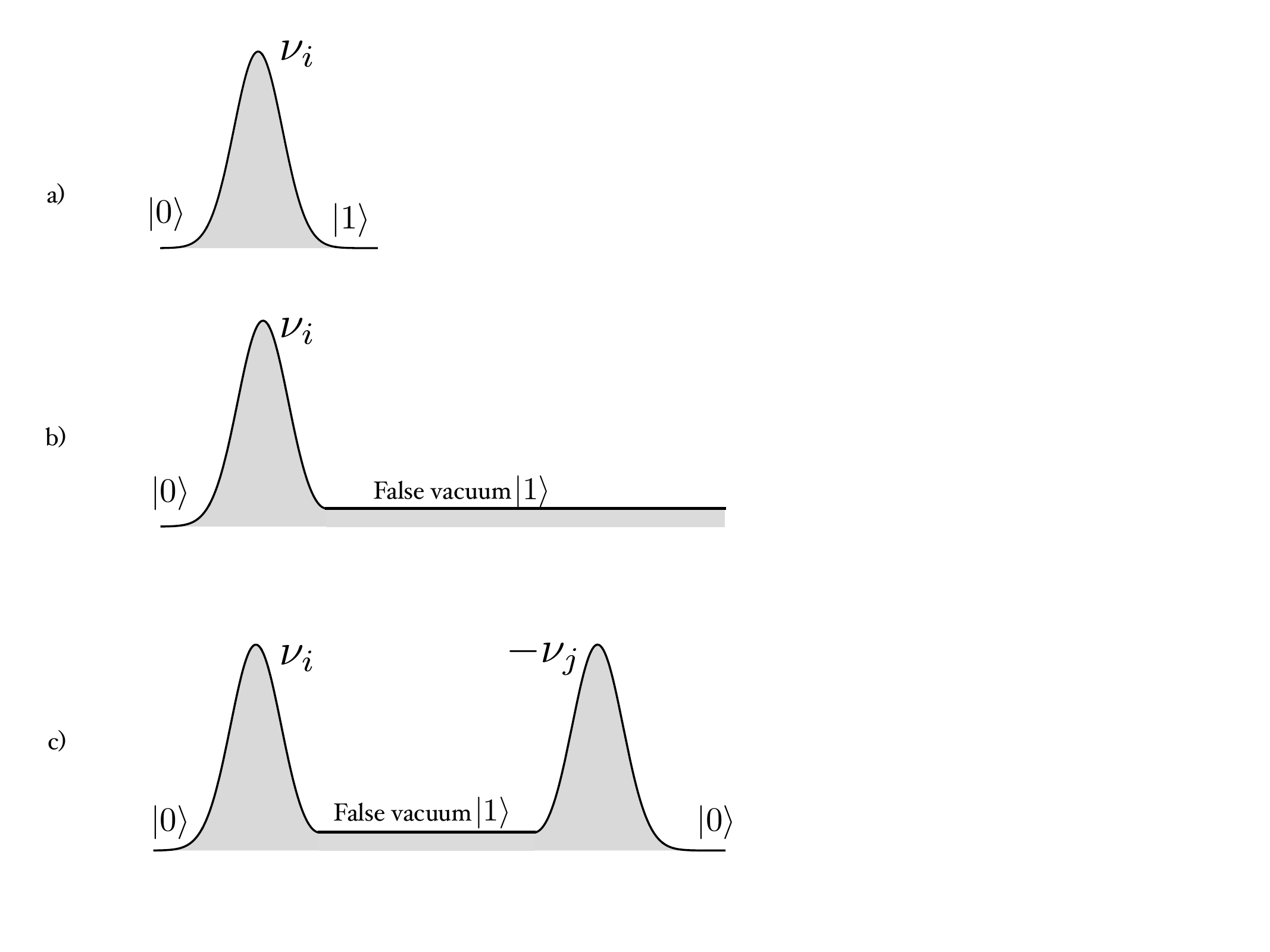}
 \vspace*{-1.5cm}
\caption{a) The BPS soliton energy density in the supersymmetric theory. The total energy of these solitons is the mass of the elementary quanta in the original sigma model. b) With small mass perturbation, one of the adjacent vacua is slightly lifted, and the would-be soliton costs infinite energy because half of the system lies in a false vacuum. Such excitations cannot exist in isolation and must be confined.  c) A soliton--antisoliton pair is a stable finite energy configuration, and transforms in a genuine representation of $\PSU(N)$. The soliton in the dual theory is the lightest massive state in the original description of the sigma model. Its mass is the mass gap of the system.}
\label{fig:Solitonabc}
\end{center}
\end{figure}

\subsection{Spectrum of the softly broken theory} 
Assume supersymmetry is softly broken by a fermion mass term. 
Then the degeneracy between the $N$ vacua should be lifted. 
The formerly degenerate vacua turn into quasivacua and their energy densities acquire fractional theta angle dependence as indicated in \eqref{branches}. 
Without loss of generality, assume $\theta =0$. 
Then the vacuum branch is at $k=0$, and  $k=1$ is the branch just above it. 
The energy density gap between the two is given by
\begin{align}
    T_1 \equiv {\cal E}_1 -{\cal E}_0=  4  m \Lambda  N  \sin^2 \frac{  \pi  }{N}=  2  m M_{1}    \sin \frac{\pi}{N}.
    \label{energydiff}
\end{align}
Even for arbitrarily small $m$, the would-be soliton solution interpolating between the vacuum $|k=0 \rangle$ to the quasi-vacuum $|k=1 \rangle$ will cost an infinite energy,
\begin{align}
    M_{1}^{\text{softly  broken}} = \lim_{\Delta x_1 \rightarrow \infty} [ M_{1} +  {\cal E}_{10} \Delta x_1] =\infty,
\end{align}
because one end of it is in the false-vacuum $|k=1 \rangle$. 
This divergence is actually physical. 
In fact, $T_1$ is nothing but the tension of the fundamental confining string attached to the elementary quanta. 
The supersymmetric theory is gapped but does not confine.  
When the supersymmetry is broken, one expects confinement, and as a result, a single $\phi$  quantum must cost infinite energy due to a semi-infinite flux tube attached to it.  

However, there are also a finite energy configurations, which may be viewed as soliton--antisoliton pairs, 
interpolating from $|0\rangle$ to $|1\rangle$ and then back to $|0\rangle$ as follows:
\begin{align}
&    |0 \rangle 
 \xlongrightarrow[]{\nu_i}  \underbrace{ |1 \rangle}_{\rm false \;  vac.} \xlongrightarrow[]{- \nu_j}    |0 \rangle. 
\end{align}
These configuration, in an approximate form, can be written as:
\begin{align}
y^{ (\nu_i, -\nu_j)} (x) = y^{(\nu_i)} (x+ \Delta x/2) +  y^{(-\nu_j)} (x- \Delta x/2)
\end{align}
where $ \Delta x$ is the characteristic size of the pair to be determined.  
For such a pair, since the width $ \Delta x$ spent  in   false vacuum is  finite, this would lead to a finite energy configuration.  

In the absence of the fermi mass perturbation, a soliton in the dual LG theory (or massive quantum in the original theory) interacts with an antisoliton via an exponentially decaying Yukawa interaction  $V(r) \sim \pm M_{1} e^{-M_{1} r}$. 
This is a natural consequence of the mass gap and the absence of confinement. 
When the fermi mass perturbation is added, since the intermediate vacuum is a false one, it costs an energy $ {\cal E}_{10} r $ to be there, and  the interaction is  modified according to 
\begin{align}
    V_{\phi_i, \phi_j^*}(r)  \sim \pm  M_{1} e^{-M_{1} r}  +  T_1 r,  
\end{align}
Here, the sign preceding the Yukawa interaction is $+$ for $i \neq j $ and $-$ for $i \neq j$. 
For $i \neq j $, the Yukawa part is repulsive and the linearly confining potential is attractive. 
The charactersitic size of the pair can be found by using $dV_{\phi_i, \phi_j^*} (r)/dr=0$: 
\begin{align}
    r_* \sim  M_{1}^{-1} \log \frac{ M_{1}}{2m \sin \frac{  \pi  }{N} } \sim  \Lambda^{-1}  \log \frac{ N \Lambda}{m}.
    \label{rstar}
\end{align}
The mass gap in the mass deformed theory is approximately equal to two times mass of the BPS soliton plus the energy stored in the flux tube holding together the soliton--antisoliton pair. 
Since the size of the pair at the static equilibrium is given by \eqref{rstar}
\begin{align}
    M_g=  2M_1 + 2 m \sin \frac{\pi}{N}  \log \biggl(  \frac{M_1}{ 2m  \sin \frac{\pi}{N} } \biggr)
    \label{massgapbos}
\end{align}

In the supersymmetric theory, the soliton of the dual formulation corresponds to the massive elementary quanta  in the fundamental ($\Box$) representation of original formulation.  
It is not difficult to realize that the above described soliton--antisoliton pair is nothing but 
\begin{align}
    \Box \otimes \overline  \Box = (\rm Adj) \oplus 1
\end{align}
These can be associated with the states created by  
the operators 
 \begin{align}
 \phi_i^* (x) e^{ i \int_x^y a}  \phi_j(y), \qquad   \sum_{i=1}^N \phi_i^*(x) e^{ i \int_x^y a }\phi_{i}(y) 
 \label{states}
 \end{align}
in the original description of the theory. 
A single  $\phi_i^*$  quanta dressed with the semi-infinite Wilson line,  $ \phi_i^* (x) e^{ i \int_x^\infty a}$ has infinite energy  due to flux tube attached and is not part of the spectrum.
This is a manifestation of linear confinement.  

Fig.~\ref{fig:SAS} shows the soliton--antisoliton pairs in the dual formulation. 
As described above, these are in precise correspondence with physical states in the Hilbert space of the original formulation.  
The mapping is:
\begin{align}
  \phi_i^* (x) e^{ i \int_x^y a}  \phi_j(y) -  \delta_{ij} \frac{1}{N}   \sum_{k=1}^N \phi_k^*(x) e^{ i \int_x^y a }\phi_{k}(y)  &\Longleftrightarrow | \nu_i, - \nu_j \rangle -  \delta_{ij} \frac{1}{N} \sum_k  | \nu_k, - \nu_k \rangle, \cr
   \frac{1}{N}  \sum_{k=1}^N \phi_k^*(x) e^{ i \int_x^y a }\phi_{k}(y) &\Longleftrightarrow  \frac{1}{N} \sum_k  | \nu_k, - \nu_k \rangle.
\end{align}

Now, let us  discuss symmetries.  The original theory is supposed to have an $\SU(N)/\ZZ_N$ symmetry, yet the dual only manifests a $\U(1)^{N-1} \times S_N$.  
In the supersymmetric theory, because of deconfinement, the physical states transform under representations of $\SU(N)$, e.g. $\Box$ and $\overline \Box$.   
However, in the mass perturbed or bosonic theory, because of confinement, the Hilbert space must be a representation of $\SU(N)/\ZZ_N$. 
Indeed, the states we found 
$ \phi_i^*    \phi_j, \phi_i^*(x) e^{\int_x^y a }\phi_{i}(y)$ generate the adjoint and singlet representations.

Let us note that the fact that our dual formulation only manifests a  $\U(1)^{N-1} \times S_N$ subgroup of  $\SU(N)$ is not an error but a common feature in two dimensions. 
A prototypical example is the duality between the massive Schwinger model with $\SU(N)/\ZZ_N$ symmetry, and its abelian bosonization which manifests only $\U(1)^{N-1} \times S_N$ symmetry.  
On the other hand, the spectrum of solitons manifests the $\SU(N)$ global symmetry for the supersymmetric case and $\SU(N)/\ZZ_N$ symmetry in the nonsupersymmetric case. 

\subsection{Remarks on the large \texorpdfstring{$N$}{N} limit } 

One of the most significant points of our analysis is that, in contrast to the claims of Ref. \cite{Witten:1978bc}, there is in fact no conflict between instantons and large $N$. 
Naively, instanton effects are order $e^{-N}$, but as our analysis shows, they can reproduce $O(N^0)$ or $O(1/N)$ effects.\footnote{Polyakov in his book \cite{Polyakov:1987ez} asserts that viewing instanton effects at large-$N$ as $e^{-N}$ is naive and wrong, and proposes some sort of plasma replacing it. In a certain sense, our refined instanton gas (with peculiar real and imaginary charges) is a concrete realization  of his view.} 
To illustrate this point, let us consider the fundamental string tension in the mass perturbed theory. 
We claim that our formula \eqref{energydiff}, which was obtained by instanton analysis, can be smoothly continued to that of the large $N$ bosonic theory. 
To start, let us quote a result of D'Adda, Luscher, and Di Vecchia \cite{DAdda:1978dle}, who have performed the large $N$ analysis of the mass perturbed theory and found the fundamental string tension at large $N$ to be
\begin{align}
    T_1 
    = \frac{12 \pi^2 \Lambda^2}{N} \frac{(1 + x)^2 \log(1+x) }{3 + \pi (2+(1+x)^2)\log(1+x)} \qquad (x \equiv m/\Lambda)
    \label{eq:DAdda-tension}
\end{align}
where there is no restriction on the magnitude of $x = m/\Lambda$. 
To reach the purely bosonic theory, we decouple the fermions by sending $x \to \infty$.
The string tension then approaches the value \cite{DAdda:1978vbw, Witten:1978bc}
\begin{align}
    T_1 \to 12 \pi \Lambda^2 /N.
\end{align}
On the other hand, taking the limit $x \to 0$, or equivalently $m\to 0$, gives us
\begin{align}
    T_1 \to 4 \pi^2 \Lambda m /N. 
\end{align}
This agrees exactly with our result, Eq. \eqref{energydiff}, which we had established via instanton analysis for $m \ll \Lambda$ but \emph{arbitrary} $N$. 
We therefore see that we can pass from the softly broken supersymmetric theory to the large $N$ bosonic theory in a completely smooth fashion. 
Indeed, we may first take $N$ large at fixed small $m$ where our formula \eqref{energydiff} is valid. From there, we can match with the formula of D'Adda et al., Eq. \eqref{eq:DAdda-tension}, and thence send $m \to \infty$. 

In the supersymmetric theory, the  mass gap has a  finite limit as $N \rightarrow \infty$, and is given by  $M_1= 2 \pi \Lambda$. In the softly broken deformation, since the lightest states are given by \eqref{states} in the adjoint representation, 
taking the large $N$ limit of \eqref{massgapbos}, we find 
\begin{align}
    M_g=  2M_1 +  \frac{2 \pi m} {N}  \log \biggl(  \frac{M_1 N }{2 \pi m } \biggr).
\end{align}
Since the correction is $1/N$ effect, the mass gap in the bosonic theory is two times the mass gap of the supersymmetric theory. 
Note that the characteristic size  of the bound state \eqref{rstar} grows loarithmically with $N$, as  
\begin{align}
r_* \sim \Lambda^{-1} \log \frac{ N \Lambda}{m} \; .
\end{align}
This follows from the fact that the linearly growing potential between $\phi_i^*$  and $\phi_j$, given by the product of the tension of the string confining the two solitons with separation, is a $1/N$ effect. 
The large $N$ limit of the tension and confining potential is 
\begin{align}
       V_{\phi_i, \phi_j^*}(r) = T_1 r  \sim  \frac{4 \pi^2 m \Lambda}{N} r
\end{align}
This is consistent with Witten's result in \cite{Witten:1978bc}. 
It implies that confining potential  in the softly broken supersymmetric theory as well as purely bosonic theory  appears at order $\frac{1}{N}$. 
In particular, at strict $N=\infty$ limit, the non-supersymmetric theory does not confine either, just like supersymmetric theory.

\section{The realization of the permutation group \texorpdfstring{$S_N$}{SN} in  
\texorpdfstring{${\cal N}=1$}{N=1} SYM on \texorpdfstring{$\RR^3 \times \SS^1$}{R3xS1} vs. \texorpdfstring{$\CP^{N-1}$}{CPN-1} model on \texorpdfstring{$\RR^2$}{R2}}

There is a well-known relation  between  ${\cal N}=1$ supersymmetric  gauge theories in four dimensions 
(e.g.  ${\cal N}=1$ SYM  theory on $\RR^3 \times \SS^1$)  and 
${\cal N}=(2,2)$  sigma models (e.g. 
 $\CP^{N-1}$ model) in two dimensions.  
In this correspondence, the chiral ring of the gauge theory gets identified with the chiral ring of the sigma model. 
The dual of gauge theory on $\RR^3 \times \SS^1$ and the LG dual of the sigma model on $\RR^2$ are both described by superpotentials of the form  
\begin{align}
{W} (Y) \,= 
\mu^a ( e^{- Y_1 } + \ldots   +  e^{- Y_{N-1} }
 + e^{-t} e^{ Y_1 + \ldots +Y_{N-1}} )
 \label{SFT-2}
\end{align}
where $e^{-t(\mu)} =  e^{-I (\mu) + i \theta}$ is the instanton factor. 
The power of cut-off scale, $a$,  is  $a=1$ for the sigma model and $a=3$ for the gauge theory. 
Solving for the vacuum in both cases, one finds $N$ isolated vacua, which are  given by 
$  e^{-Y_{i}^*} = e^{-t/N}  e^{i  \frac{2 \pi k}{N}}$.  
These are associated with the sectors of spontaneously broken discrete chiral symmetry $\ZZ_{2N} \rightarrow \ZZ_2$. 
Both theories possess  $ \tr (-1)^F = N$ vacua, and there are domain walls interpolating between them. 
Using the dimensional transmutation,  $\mu^a    e^{-t/N} = \Lambda^a$, where $\Lambda$ is the renormalization group invariant strong scale, and denoting the fluctuations around these vacua by $\widetilde Y_i =   Y_i-  \frac{t}{N}$, one ends up with 
\begin{align}
{W} ( \widetilde Y) \,= 
\Lambda^a ( e^{-  \widetilde Y_1 } + \ldots   +  e^{-  \widetilde Y_{N-1} }
 +  e^{ \widetilde  Y_1 + \ldots +  \widetilde Y_{N-1}} )
 \label{SFT-3}
\end{align}
In the supersymmetry literature, the superpotentials for these two theories seem to be identified as supersymmetric affine Toda theory

However,  our construction suggests that the Euclidean statistical field theory description of these theories are actually different in a crucial way.  
In particular, the  explicit superpotential we find for the sigma model and for the gauge theory are given by 
\begin{align}
{W}  \, &= \, \Lambda \,\sum_{i=1}^N e^{- {\bm \nu}_i. \bm Y } \qquad {\rm for}  \;\;  {\cal N}=(2,2) \; \CP^{N-1},  \cr
{W}  \, &= \, \Lambda^3 \,\sum_{i=1}^N e^{- {\bm \alpha}_i. \bm Y } 
 \,\;\quad {\rm for}   \;\;   {\cal N}=1 \; {\rm SYM},
 \label{SFT-compare}
\end{align}
where the $\bm \nu_i$ are the weights of the defining representation and the $\bm \alpha_i$ are the simple roots of $\SU(N)$. 

In particular, in the sigma model, the system has an $\SU(N)/\ZZ_N$ global symmetry, and  \eqref{SFT-2} gives a description of the system in which only the subgroup $\U(1)^{N-1} \rtimes S_N$ is realized manifestly. 
The important point we wish emphasize is that the Weyl group is an exact symmetry of the description and is unbroken.  
In the gauge theory on small $\RR^3 \times \SS^1$, the microscopic theory has an $\SU(N)$ gauge symmetry, and  \eqref{SFT-2} gives a description of the system in which 
$\SU(N)$  is Higgsed down to the subgroup $\U(1)^{N-1} \rtimes \ZZ_N$, where $\ZZ_N$ is the cyclic subgroup of $S_N$. 
The important point is that in this description, the non-abelian discrete permutation group $S_N$ is Higgsed (or broken). 
\begin{align}
    S_N \rightarrow  \left\{ \begin{array}{l}
    {\rm Unbroken \;  in} \qquad {\cal N}=(2,2) \; \CP^{N-1} \;\; {\rm on}  \;\; \RR^2   \\
   {\rm Higgsed  \;  in} \qquad \mathbb  {\cal N}=1 \; {\rm SYM} \;\; {\rm on \; small}  \;\; \RR^3 \times \SS^1
     \end{array} \right.
\end{align}

This has many physical implications. 
In ${\cal N}=1 \; {\rm SYM} \;\; {\rm on \; small}  \;\; \RR^3 \times \SS^1$, the lowest lying  part of the spectrum is  $(N-1)$ dual photons and  they 
are not  degenerate.   (Degeneracy factors for the lowest $N-1$ states has either $(2 \ldots 2)$ pattern or  $(2 \ldots 21)$
pattern.)   On the other hand, in  the dual Landau--Ginzburg description of the sigma model, 
the spectrum  is the standard $N-1$ dimensional   irreducible representation $D_{\rm std}$ of $S_N$: 
 \begin{align} 
  m_p& = \Lambda ( \Lambda L N)^2 \sin^2\left( \frac{  p \pi}{N}  \right), \qquad p=1, \ldots, N-1  \qquad   \; {\rm SYM} \cr
 m&= \Lambda, \;\qquad  D_{\rm std}   \in S_N,  \; \;  {\rm dim}(  D_{\rm std} )= N-1,   \qquad  {\rm LG} \;  {\rm on} \;  \RR^2 
 \end{align}
 
In gauge theory on small $\RR^3 \times \SS^1$, the monopole-instantons (denoted by $e^{- {\bm \alpha}_i. \bm Y }$  operators)  are finite size physical configurations, and they can be viewed as the genuine physical fractional constituents of a 4D instanton, as a consequence of adjoint Higgsing, which breaks the permutation group.

By contrast, in the sigma model on $\RR^2$, the generalized vortices (denoted by $e^{- {\bm \nu}_i. \bm Y }$  operators) do not correspond to a physical fractionalization of the instanton due to absence of adjoint Higgsing. 
Rather, the collection of these vortices give a parametrization of the 2d instanton. However, on compactifying the $\CP^{N-1}$ model on small $\RR \times \SS^1$ with a symmetry-twisted boundary condition, the instanton physically fractionates into $N$ constituents. The boundary condition explicitly breaks the permutation symmetry and as a result, the fractional instantons are characterized by the roots $\bm \alpha_i$ rather than the weights $\bm \nu_i$, as in the gauge theory on $\RR^3 \times \SS^1$.


\section{Remarks on literature}
Some elements of our construction have appeared in earlier literature, but not the full statistical field theory, nor the Gibbs distribution we propose for the subsystem. 
Below, we highlight the works that have been most influential on our analysis and state clearly our points of disagreement. 

The classical interactions between the instantons in the $\CP^1$ model were first calculated 
by Gross \cite{Gross:1977wu}, following a  similar calculation by Callan, Dashen, and Gross \cite{Callan:1977gz} in 4D QCD. 
The generalization to $\CP^{N-1}$ was derived in underappreciated work by Bardakci, Caldi, and Neuberger \cite{Bardakci:1980pj}. 
Recall that the classical instanton--anti-instanton interaction is non-zero and of dipole-dipole type while the classical instanton-instanton is
zero. 
Although both works attempted to construct a statistical field theory, neither of them realized the necessity of introducing both real and imaginary charges. 
This makes their analysis inconsistent with classical data, that is, inconsistent with the structure of the classical moduli space of instantons.  
Furthermore, Gross attempted to formulate an instanton as genuinely two split configurations, called merons. 
But as is by now clear, there is no physical fractionalization of an instanton into fractional action configurations under the given circumstences.  
The instanton is a single lump. 

The necessity of  both real and imaginary charges to produce $V_{\cal I \cal I}  = 0$  and 
$  V_{\cal I \overline{\cal I}} \neq 0 $  simultaneously is due to Forster in  $\CP^{1} $ model \cite{Forster:1977jv}.  
In our opinion, this was an important step.  
However, the statistical field theory Forster constructs is not consistent with the fact that classically, instantons come in all sizes. 
In particular, the Gibbs distribution he constructs is similar to a theory with fixed-size instantons (fixed size generalized permanent dipoles) shown in right panel of Fig.\ref{fig:plasma1}. 
As a result, Forster's statistical field theory does not capture the vacuum properties correctly even at the classical level. 
  
The idea that one can associate charges to moduli parameters is due to  Fateev,  Frolov, and Schwarz  \cite{Fateev:1979dc} and  Berg and L\"uscher  \cite{Berg:1979uq}. 
These works calculate the fluctuation determinant around multi-instanton configurations in the  $\CP^{1}$ case correctly, and  determine the quantum-induced  Coulomb interactions $O(g^0)$  between the constituents of an instanton. 
They then build an effective theory based \emph{only} on the proliferation of instantons.  
This is a theory of something, but it does not describe the vacuum structure of the  $\CP^{1}$ model. 
If one wishes to learn about the vacuum structure of the $\CP^{1}$ model, it is fundamentally inconsistent to drop the classical  $O(1/g^2)$  interactions  and base a theory on quantum-induced ones.   
Dropping anti-instantons amounts to that.  
Furthermore, the two-body quantum interaction is an accident for $\CP^1$ -- for general $\CP^{N-1}$, the quantum-induced interactions are $N$-body, while the classical interactions between constituents are always two-body. 
Bukhvostov and Lipatov \cite{Bukhvostov:1980sn} take into account both the classical and quantum interactions between the constituents in the $\CP^{1}$ model, aiming to generalize Fateev et al. to systems with both instantons and anti-instantons. (See also \cite{Bazhanov:2016glt,Bazhanov:2017xky})   
The effective theory they suggest is a standard two component Coulomb gas with real charges, based on the assumption that the magnitude of quantum interactions overwhelms the classical ones.  
But in controllable theories such as the supersymmetric $\CP^{1}$ model or the soft mass deformation thereof, the situation is the reverse. 
Had these authors assumed that the classical interactions dominate, they would have ended up with imaginary charges and thence an unstable theory at the first order in the fugacity expansion. 
  
Finally, the work that we found most beneficial but ironically, the one we disagree with most is the one by Witten  \cite{Witten:1978bc}.  
We find the large-$N$  analysis in this paper (as well as \cite{DAdda:1978vbw, DAdda:1978dle}) inspirational. 
However, Ref.~\cite{Witten:1978bc} argues that instanton effects vanish exponentially as $e^{-cN}$ for large $N$. 
For example, he argues that the $\theta$ dependence according to instanton analysis must be of order $e^{-cN}$, so that the $\theta$  dependence that appears at $O(1/N)$ according to large-$N$ analysis is unlikely to arise from instantons. 
In our work, we were able to derive the large-$N$ results from a refined instanton analysis. 
Therefore, our work and \cite{Witten:1978bc} must disagree  about instantons at a fundamental level.  
In what consideration do we differ? 
In particular, we believe that Ref.\cite{Witten:1978bc} did not attempt to provide 
the Gibbs distribution (in which instanton size is \emph{classically} a modulus) and does not 
consider the \emph{classical} interactions between instantons and anti-instantons. 
As we showed in Fig.~\ref{fig:plasma1} (left), once this is done, one realizes that one must construct the Gibbs ensemble for subsystems  based on charges attached to moduli. 
The resulting statistical field theory correctly produces $O(1/N)$ effects, fractional theta dependence, and other physical properties that arise from the large-$N$ analysis resolving potential inconsistencies discussed by Witten. 

\section{Conclusions and outlook}

Instantons in 4D gauge theories and 2D sigma models have been known now for almost half a century.
There have been many works on the possible description of the vacuum structure of these quantum field theories based on instantons. 
Despite the fact that instanton methods are extremely successful for theories in which  instanton have a natural characteristic size, such as spontaneously broken gauge theories and  quantum mechanics, for unbroken asymptotically free gauge theories, the standard instanton approach aiming to understand the vacuum structure seems to be a failure, not capturing the most interesting aspects of the nonperturbative physics. 

In this work, we have argued that the existing standard instanton analysis is not the right implementation of the instanton idea in asymptotically free quantum field theories, and have introduced a refined analysis. 
In particular, we have stressed the importance of a formulation in which one respects all classical properties of the instantons: the classical interactions between ${\cal I}$ and $\overline {\cal I}$ and absence thereof between  ${\cal I}$ and $ {\cal I}$, and the structure of the classical moduli space.  
At the level of classical analysis, the coupling constant can be kept constant and set to a cut-off value. 
The running of the coupling and dimensional transmutation enter at a later stage. 
The classical interactions and consistency conditions allow us to express an instanton in terms of $N$ moduli parameters, with real and imaginary charges  $(i {\bm \nu}_i,   {\bm \nu}_i ) \in i \weightlattice \oplus  \weightlattice$.  
Then, the existence of the classical moduli space (or the fact that instanton can come in any size and orientation at the classical level) implies that when we build a Gibbs distribution for subsystems, it is the constituents that enter into the statistical sum, not merely full instantons themselves. 
This is depicted in Fig. \ref{fig:plasma1}(left). 
This is the major difference between a theory in which the instantons are of fixed size and a theory in which instanton size (and orientation) can be classically anything.  
We believe that this crucial step, which is at the very heart of the classical refined instanton analysis, has been missing in all past analyses.

Our construction gives a successful description of the vacuum. 
It resolves many long-standing conflicts between instantons and reliable large $N$ analysis.  
Moreover, the application of our methods to the supersymmetric ${\cal N}=(2,2) $ 
$\CP^{N-1}$  sigma model reproduces the mirror dual description based on generalized vortices. 
We find both of these features quite remarkable.

Of course, one would ultimately hope to achieve a similar understanding of the instanton gas in four-dimensional asymptotically free gauge theories. 
The success of our perspective in the 2D $\CP^{N-1}$ model makes this hope seem viable. 
Indeed, there are many similarities between the 4D gauge theory instantons and the 2D sigma model instantons considered here.
For instance, the moduli space of the 4D $\SU(N)$ instanton has dimensionality $4N$, which makes plausible the possibility of describing the instanton as a collection of $N$ pointlike constituents. 
Also, the classical $\inst \inst$ interaction vanishes exactly, while the $\inst \antiinst$ interaction is of dipole--dipole type.
Unfortunately, the electrostatic analogy, which was so useful in the 2D case, breaks down in the 4D case, because the dipole--dipole interaction of $\inst \antiinst$ is that of \emph{magnetic} dipoles, and it is not straightforward (nor even clear if possible) to reproduce this from a description of the instanton as a collection of pointlike constituents.
It is an important and nontrivial task to construct a formalism which captures these classical data correctly. 
Nevertheless, we do not think that these difficulties should be impossible to overcome. 
We leave this for future work.

\newpage 


\acknowledgments
We thank David Gross for useful discussions. 
The work of M.~\"U. was supported by U.S. Department of Energy, Office of Science, Office of Nuclear Physics under Award Number DE-FG02-03ER41260. 

\appendix

\section{Overview of  mirror symmetry}\label{sec:introduction}
In this appendix, we provide an overview of the mirror symmetry between the supersymmetric  $\CP^{N-1}$ sigma model and 
and Landau--Ginzburg theory of Toda type from a physical perspective. 
Our description follows very closely Hori and Vafa's original proof \cite{Hori:2000kt}.  
Following them, we start by embedding the nonlinear sigma model into the gauged linear sigma model (GLSM). 
Differing from them, we rely much less on supersymmetry. 
In subsection.~\ref{AHM},  we provide  a  careful analysis of nonperturbative effects and abelian duality in the nonsupersymmetric abelian Higgs model.\footnote{ Although this topic is a textbook material by now, the standard treatments  ignore the instanton interactions (since they are short ranged) and do not build an effective field theory on them \cite{Coleman:1985rnk, Deligne:1999qp}. 
Ref. \cite{Komargodski:2017dmc} takes into account one type of interaction between the instanton, induced by $\vartheta$  and  ignores the ones associated with $\varrho$
in \eqref{vortexAHM}. 
Keeping both fields in the dual description is more useful. 
Then the relation to the supersymmetric abelian Higgs model and mirror dual \cite{Hori:2000kt} become  transparent.} 
This analysis is also useful for our refined instanton analysis which uses a generalized  Coulomb gas with real and imaginary charges.  
In the main text of this paper, we show that our refined instanton analysis provides an alternative derivation of the mirror dual of the supersymmetric sigma model. 

Consider ${\cal N}=(2,2)$ supersymmetric $\U(1)$ gauge theory with $N$ chiral superfields. 
The content of the theory is a $\U(1)$ vector multiplet $V \sim (\sigma, A_{\mu}, \lambda_{\pm}, \overline  \lambda_{\pm})$, whose field strength is denoted $\Sigma$ which is a chiral twisted superfield, and $N$  chiral matter fields $\Phi_i \sim (\phi_i, \psi_{i\pm} )\  (i=1, \ldots, N$),  whose charges are $Q_i=+1$.  
The Lagrangian of the gauged linear sigma model (GLSM) in superfield notation is given by 
\begin{align}
L=\int d^4\theta
\left(\,
\sum_{i=1}^N \overline \Phi_i\,e^{2Q_iV}\Phi_i
-{1\over 2e^2}  \overline \Sigma\Sigma
\,\right)
+{1\over 2}\Biggl(
-\int d^2\widetilde{\theta}~t\,\Sigma
\,+\, c.c.\,
\Biggr),
\label{lag}
\end{align} 
The theory has three parameters, $e^2$ with dimensions of mass, and a dimensionless complex parameter $t= r-i \theta$, where $\theta$ denotes the topological theta angle of the $\U(1)$ theory: 
\begin{align}
[e^2]= +2,  \quad  [t]= [r-i \theta] =0. 
\end{align}
 
Classically, the global symmetry of the theory  is  $\SU(N)/\ZZ_N$ where $\ZZ_N$ is a part of $\U(1)$ gauge redundancy, and needs to be factored out. 
The theory also possesses a $\ZZ_{2N}$ axial symmetry. 
In components, the action in Euclidean space can be expressed as
\begin{eqnarray}
S_E\,= \,{1\over 2\pi}\int d^2x&& 
\Bigl(~
|D_{\mu}\phi_i|^2+|\sigma\phi_i|^2
+{1\over 2e^2}|\partial_{\mu}\sigma|^2+ D |\phi_i|^2 +  {1\over 2e^2}(F_{12}^2+D^2)
-r_0 D +i\theta F_{12}
\nonumber\\
&&\!\!
-2i \overline \psi_{i-} D_{\bar z}\psi_{i-} + 2i \overline \psi_{i+} D_z\psi_{i+}
+\overline \psi_{i-}\sigma\psi_{i+}+ \overline \psi_{i+}\overline{\sigma}\psi_{i-}
\nonumber\\
&&\!\!
+{1\over e^2}\left(\,-i\overline \lambda_-\partial_{\bar z}\lambda_-
+i \overline \lambda_+\partial_z\lambda_+\,\right)
\nonumber\\
&&\!\!
+i\left(\,
\phi^{\dag}_i \lambda_-\psi_{i+}-\phi^{\dag}_i \lambda_+\psi_{i-}
- \overline \psi_{i+} \overline \lambda_-\phi_i+  \overline \psi_{i-}\overline \lambda_+\phi_i\,\right)~\Bigr).
\label{SE}
\end{eqnarray} 
The repeated indices here are summed over.   

The classical potential, which determines the classical supersymmetric vacua, can be obtained by solving the equation for the  D-term,  $D=-e^2  \left(\sum_{i=1}^N |\phi_i|^2-r_0 \right) $ and is given by 
\begin{align}
U= 
 \sum_{i=1}^{N}  |\sigma\phi_i|^2  +  {e^2\over 2}
\biggl(\,\sum_{i=1}^N |\phi_i|^2-r_0\,\biggr)^{2}
\label{pot-vac}
\end{align}
The classical minima are given by the configurations for which the potential vanishes.  
One obtains 
\begin{align}
\sigma =0, \quad  \sum_{i=1}^N |\phi_i|^2=r_0.
\label{vev}
\end{align}
The set of minima modulo $\U(1)$ gauge redundancy describes the classical vacuum manifold. 
This is $N-1$ complex dimensional complex projective space: 
\begin{align}
 \CP^{N-1} = \Big\{  (\phi_1, \ldots, \phi_N) \Big| \sum_{i=1}^N |\phi_i|^2= r_0 \Big\}/ \U(1) 
 \label{cvm}
\end{align}
Classically, any choice of a point on the vacuum manifold breaks the global $\SU(N)/\ZZ_N$  symmetry down to $\U(N-1)/\ZZ_N$.  
The accompanying massless bosons would be the Nambu--Goldstone (NG) particles. 
For example, for $N=2$ or  $\CP^{1}$, the global $\SU(2)/\ZZ_2 = \SO(3)$ symmetry should be broken down to  $\U(1)$, and there would be two massless particles.  
Goldstone theorem states that the implications of this classical analysis persist quantum mechanically for $d \geq 3$, but in $d= 2$ dimensions, the story is different. 
There is no spontaneous breaking of the continuous 0-form symmetries in $d=1+1$ and no corresponding NG bosons exist. 
In other words, a scalar does not freeze at a point, rather it explores the full geometry of the vacuum manifold. 
Because of this, the low energy model for $N\geq 2$ reduces to a nonlinear sigma model on complex projective space  $\CP^{N-1}$.

\vspace{0.3cm}
The perturbative spectrum of the $N$-component  model has a sharp difference between $N=1$  and $N \geq 2$ cases. By supersymmetric Higgs mechanism, 
all the modes in the vector multiplet $(v_\mu, \sigma, \lambda_+,   \lambda_-)$ acquire  a mass $m= e \sqrt {2r_0}$. 
Out of $N$  $\phi_i$, 
the modes tangential to  $\sum_{i=1}^N |\phi_i|^2= r_0$ remain gapless 
perturbatively, accounting for $N-1$ complex degrees of freedom.  Locally, one real mode in $\phi_i$ acquires mass and one mode is  eaten by the gauge field  $v_\mu$ to render the gauge field massive.  
 Similarly,  the fermion mode  for which   $\phi_i^{\dagger} \psi_{i\pm} \neq 0 $ pairs up with $\lambda^{\mp}$ rendering both massive and the remaining 
$N-1$ fermions obeying   $ \sum_{i=1}^N \phi_i^{\dagger} \psi_{i\pm}= 0 $ remain massless to all orders in perturbation theory.  
\begin{align}
&N=1: \qquad {\rm All \;  dof \;  are \;  gapped \; perturbatively.} \cr
&N\geq2  \qquad \;\;\; {N-1 }\;  {\rm IR \;  modes \;  remain \;  gapless \;  perturbatively. }
\end{align}
The remaining $N-1$ chiral multiplets  is described in terms of  a non-linear sigma model on complex projective space  $\CP^{N-1}$. This is an asymptotically free theory that becomes strongly coupled in the infrared. Our goal is to figure out non-perturbative dynamics both in non-supersymmetric and supersymmetric theory.

\subsection{Abelian Higgs model: \texorpdfstring{$N=1$}{N=1} bosonic case} 
\label{AHM}

In this section, we provide a detailed description of the vacuum structure of the abelian Higgs model, in terms of 
the instantons  (vortices), their classical interactions and  their statistical field theory.   From the statistical  field theory, in both the bosonic and supersymmetric cases,  we can read off  non-perturbative phenomena such as confinement  (or absence thereof), chiral symmetry breaking, theta angle dependence etc.  

\vspace{0.3cm}
Consider $\U(1)$ gauge theory coupled to  complex scalar $\phi$, non-supersymmetric version of \eqref{SE} with $N=1$. The Lagrangian is given by 
\begin{equation}
    \mathcal{L}=|D_\mu \phi|^2+\lambda(|\phi|^2-r_0)^2  +{1\over 2e^2}|F_{12}|^2 -{i \theta\over 2\pi}F_{12}, 
    \label{eq:Lagrangian_AbelianHiggs}
\end{equation}
where we introduced $\lambda$ as the coefficient of the bosonic potential. 

\vspace{0.3cm}
For $N=1$, every particle in the spectrum    is gapped out due to Higgs mechanism.  First, note that the vacuum manifold is given by  $|\phi|^2 =r_0$ modulo $\U(1)$ gauge transformation, $\SS^1/U(1)$ \eqref{cvm}, which is  a single point.   Writing $\phi =|\phi| e^{i \varphi}$,  
$|\phi|$ obtains a mass as it is normal to $\SS^1$ manifold, and   $\varphi$ is eaten by the gauge field   $v_\mu$ rendering it massive. 

\vspace{0.3cm}
To all orders in perturbation theory,  the theory is in the Higgs phase, i.e., not confining. However, non-perturbatively,  
the theory has  vortices (which are instantons in 2d).  The vortices  have the same form as in \eqref{eq:vortex-op-1}
 \begin{align}
{\cal V}  \sim  \rme{-S} e^{  - \varrho (x) +  i  \vartheta (x) + i \theta}, \qquad 
\overline  {\cal V}  \sim \rme{-S}  e^{  - \varrho (x) -  i  \vartheta (x) - i \theta} 
\label{vortexAHM}
 \end{align} 
 but their mutual  interactions  are different from \eqref{eq:IntVVbar}.  Since the  abelian Higgs model  is  fully Higgsed, the interactions become  short ranged. 
Due to perturbatively generated masses,  $m_{\varrho}, m_\vartheta$,  the vortex-vortex and vortex-anti-vortex correlators  take the form: 
 \begin{align}
\langle {\cal V}({\bf r} )   {\cal V}({\bf 0} )  \rangle     \sim  e^{  - V_{{\cal V} {\cal V}} (r)  }, \qquad 
\langle {\cal V}({\bf r} )  \overline  {\cal V}({\bf 0} )  \rangle     \sim  e^{  - V_{{\cal V} \overline  {\cal V}} (r)  },
\label{vortexAHM2}
 \end{align} 
where the short range interactions due to ${\varrho}$ and ${\vartheta}$ exchanges   are given by 
\begin{align}
V_{{\cal I} {\cal I}} (r)   &= -  \frac{1 }{2 \pi}  K_0 (m_{\varrho} r)  +   \frac{1 }{2 \pi}  K_0 (m_{\vartheta} r)     \qquad  {\rm vortex-vortex \;  interaction }  \cr
V_{{\cal I} \overline {\cal I}}  (r)   &= -  \frac{1 }{2 \pi}  K_0 (m_{\varrho}  r)  -    \frac{1 }{2 \pi}  K_0 ( m_{\vartheta} r) 
 \qquad  {\rm vortex-antivortex \;  int. } 
 \label{short-range}
\end{align}
where $K_0(m  r)$ is the modified Bessel function.    The interaction due to $\varrho$  exchange is attractive in both cases and  
the ones due to $\vartheta$  exchange is attractive  for opposite charges and repulsive for same charges. 
In the standard textbook discussions \cite{Coleman:1985rnk, Deligne:1999qp}, the ${\varrho}$ and ${\vartheta}$ are set to their mean field value and one considers proliferation of (non-interacting) vortices  with amplitudes 
${\cal V}  \sim  e^{-S+ i \theta}, \overline  {\cal V}  \sim e^{-S- i \theta}$.  If $m_\varrho \gg  m_\vartheta$,  then one can integrate out the 
$\varrho$ field and write an EFT based on $ \vartheta $ as it was done in 
Ref. \cite{Komargodski:2017dmc}. But depending on the physical circumstances, one can have $m_\varrho <  m_\vartheta$ (corresponding to type-1 superconductors, where vortices attract),    $m_\varrho >  m_\vartheta$ 
(corresponding to type-2 superconductors, where vortices repel),     $m_\varrho =  m_\vartheta$ where same type vortices do not interact. Therefore, to get a more physical picture which captures various interesting limits, it is better to keep both $m_\varrho$ and  $m_\vartheta$ fields in vortex operator  \eqref{vortexAHM}.

\vspace{0.3cm}
The limit that is relevant to the   ${\cal N}=(2,2)$ supersymmetric   theory correspond to the case where 
$m_{\varrho}= m_\vartheta$. In that case, 
\begin{align}
V_{{\cal V} {\cal V}} (r)   &= 0     \qquad    \qquad   \qquad   \;\; \;   {\rm vortex-vortex \;  interaction }  \cr
V_{{\cal V} \overline {\cal V}}  (r)   &= -  \frac{1 }{ \pi}  K_0 (m_{\vartheta}  r)   \qquad  {\rm vortex-antivortex \;  int. } 
 \label{short-range-susy}
\end{align}
As expected, vortex-vortex interaction becomes zero and the one between vortex and anti-vortex is doubled compared to the pure $\vartheta$ induced interactions.

\vspace{0.3cm}
It is also useful to check the asymptotics. 
In the large distance limit, the interaction potential decays exponentially fast, indicating its short-range nature. 
If  we ungauge the theory,  where the gauge coupling is set to zero, $e=0$,  the interactions become  long range Coulomb  interaction in 2d. Both limits can be seen from  the asymptotics of the Bessel function:
   \begin{align}
 \frac{1}{2 \pi} K_0(x)   \rightarrow \left\{  \begin{array}{ll}
-  \frac{1}{2 \pi} \log x &    \qquad    x  \rightarrow 0 \cr  \cr
\frac{1}{2 \pi} \sqrt{ \frac{ \pi }{2x } } e^{-x},   &   \qquad    x  \rightarrow \infty
\end{array}  \right.
   \end{align}
 The ungauged  limit of the interactions  has many interesting features of its own, famously appears in the BKT phase transition, and it also plays a role in our refined instanton analysis.

\vspace{0.3cm}
The Euclidean vacuum  of the bosonic theory can be described as a grand canonical ensemble of vortices, which interact via short-ranged interactions  
\eqref{short-range}. An effective Lagrangian for the corresponding statistical field theory can be written as:
\begin{align}   
 {\cal L}&=\frac{g^2}{4 \pi^2}  \left(   (\diff \varrho)^2  +  (\diff \vartheta)^2  \right)  + \frac{1}{2} m_{\varrho}^2 \varrho^2     + 
  \frac{1}{2} m_{\vartheta}^2   \vartheta^2    
- 2 \zeta    e^{- \varrho}  \cos( \vartheta   )  +  
 \ldots  \qquad |\theta|< \pi
 \label{dual-AHM}
\end{align} 
For general $\theta$ angle, the mass operator needs to be replaced with periodic function:
\begin{align}
V(\vartheta) =  \half m_{\vartheta}^2  \min_{n \in \Z} (\vartheta -  \theta- 2 \pi n)^2 
\label{cuspy}
\end{align}
It is useful to provide a simple derivation of the mass term, as similar manipulations also helps to calculate vacuum expectation values of the Wilson loops. 

\vspace{0.3cm}
First, recall that  the fields in the dual formulation of the abelian Higgs model, in particular $\vartheta$,   is uncharged under the $\U(1)$ gauge field. Hence,    the gauge field only appears through the field strength $f= \diff a$ in the Lagrangian. If we view the theory on some compact manifold $M_2$,  we can replace the  integral 
 over $a$ with integral over $f$  by summing over all integer  flux sectors labelled  by $\nu \in \Z$,  and further using Poisson resummation formula, we reach:
\begin{align} 
\int Da \rightarrow &   \int Df \sum_{\nu \in \Z} \delta \left( \nu - \frac{1}{2\pi} \int f \right) = \int Df  \sum_{n  \in \Z}  e^{ i n    \int f  }
\end{align} 
Since gauge field strength  becomes a Gaussian variable, it can be integrated out exactly, reproducing a mass term for the scalar  
$\vartheta$. 
\begin{align}
&\int Da \;   \exp{ \left[{ -{1\over 2e^2}\int |f|^2 + \frac{\im}{2 \pi}    \int \vartheta f  +  \frac{\im}{2 \pi}    \theta \int  f   } \right] }  \cr
&=      \sum_{n  \in \Z} \;     \int Df  \;   \exp \left[ { -{1\over 2e^2}\int |f|^2 + \frac{\im}{2 \pi}    \int (\vartheta  + \theta +  2 \pi n) f  
} \right]\cr
&={ \rm const.} \;   \sum_{n  \in \Z} \;   {\exp} \left[  { - \frac{1}{2} \frac{e^2 }{4 \pi^2}  \int  \left ( \vartheta +  \theta +  2 \pi n  \right)^2} 
\right]
\label{atof}
\end{align}  

\vspace{0.3cm}
In the deep infrared of the abelian Higgs model, we can forget about all  gapped fluctuations, since they freeze due to mass term. 
The partition function for the \eqref{dual-AHM}  is dominated by the mean field of the functional integral, given by $\vartheta=-\theta, \varrho=0$, leading to the theta angle dependence of the partition function and vacuum energy density  
\begin{align}
    Z(\theta)&= 
   e^{ -V E(\theta)} =    e^{ -V\left(-2K e^{-S_I}\cos {\theta} \right)}= 
   \sum_{n,\overline{n}\ge 0}{V^{n+\overline{n}}\over n! \overline{n}!}K^{n+\overline{n}}e^{-(n+\overline{n})S_I}e^{i (n-\overline{n})\theta} 
\end{align}
In the last step,  we expressed the  partition function as a grand canonical ensemble of a  non-interacting vortex gas with complex fugacity  $\zeta= e^{-S_I}   e^{ i \theta}$.    

\vspace{0.3cm}
Now, we are ready to calculate the expectation value of the Wilson loop. 
Consider the insertion of the Wilson loop with charge $q \in \R$. By using Stokes theorem,  we can rewrite it 
as an integral over the  field strength. 
\begin{align}
 W_q(C)  = 
 \exp \left[ i q \int_C a \right] = \exp  \left[  i q \int_D f \right]   =   \rme  \left[ i q \int_{M_2}  \Theta_D f   \right] 
\end{align}
where $\Theta_D ({\bf r})$ is the bump function, defined as 
\begin{align} 
\Theta_D({\bf r}) = \left\{  \begin{array}{cc} 
1, \qquad  &  {\bf r}  \in D    \\
0, \qquad  &{\bf r}  \notin D    
\end{array}
\right.
\label{bump}
\end{align}
Therefore,  insertion of the Wilson loop modifies the mass term \eqref{atof} into  
\begin{align}
V(\vartheta) \rightarrow  \half m_{\vartheta}^2  \min_{n \in \Z} (\vartheta + q  \Theta_D  +\theta+ 2 \pi n)^2  
 \end{align}
Therefore, the vacuum expectation value of the Wilson loop is equal to
  \begin{align}
    \langle W_{q}(C)\rangle &= \frac{ Z(\theta + q  \Theta_D)} {Z(\theta)}    
  \end{align} 
This result is simple and physical.  Recall that the theta angle in 2d   can be viewed as an insertion of charge $\pm \frac{ \theta } {2 \pi} $ at 
$x= \pm \infty$,  as capacitor plates at infinity, with an electric field in between. The insertion of the Wilson loop with charges $\pm q $  modifies  the field in between  into $ ( \frac{ \theta + 2 \pi q  } {2 \pi})  $. Therefore, 
   \begin{align}
  \langle W_{q}(C)\rangle &= \frac{ e^{ -\calA  E(\theta+2\pi q )- \calA_{\rm c} E(\theta)}} {   e^{ - V E(\theta)}}   
  =  e^{-   T_q A(D)}, \qquad (\rm bosonic \; case)
  \label{VEV}
  \end{align}
where $\calA $ is the area enclosed by $C$, and $\calA_{\rm c}= V-\calA$  is the area of the complement.  The string tension is just the difference of energies:
 \begin{align}
 T_q =  E(\theta + 2 \pi q) - E(\theta) =    4 e^{-S_v} \left( \cos \theta  - \cos ( \theta+ 2 \pi q) \right), \qquad {\rm for} \;\; |\theta|<\pi 
 \end{align} 
For $q \in \Z$, the string tension is zero because of the pair creation of dynamical charges and     string breaking.\footnote{ 
 Note that  in the confinement criterion, we use external  charges such that $q$ is not an integer,  i.e. it is a representation of the  the universal covering group $\R$  but not $\U(1)$. Since the  probe operators are in  $\R$  gauge group, sometimes it is stated that we work with  $\R$ gauge theory. A useful way around this is to define  a theory in which dynamical charges have integer charge 
 $p \geq 2$. This is sometimes called charge-$p$ abelian Higgs model.  Charge-$p$  model  has an exact  1-form  
  $\ZZ_p^{[1]}$ symmetry, which can be used to sharpen what one means by confinement, and to derive various mixed anomalies involving 1-form symmetry.  
  For various aspects of the charge-$p$ model,  see the   Appendix B in  \cite{Tanizaki:2022ngt}. }

\vspace{0.3cm}
The same result can also be obtain from the physical picture of the grand canonical ensemble. 
In the deep infrared,  all gapped fluctuations freeze.  
The only information that remains is the relation between the vortices  and Wilson loops, $W_q(C)$.  Let $D$ denote the surface for which  $\partial D = C$, and $ {\bf r} $ the position of vortex.  Then, classically, 
\begin{align} 
 W_q(C) = \left\{  \begin{array}{cc} 
 e^{i 2 \pi q} \qquad  &{\bf r}  \in D     \\
 1, \qquad  &  {\bf r}  \notin D    
\end{array}
\right.
\label{vortex-Wilson}
\end{align}
i.e. if the vortex is inside the loop, the holonomy gives a non-trivial phase, otherwise, the holonomy is trivial. Therefore, 
 $   \langle W_{q}(C)\rangle $ can be written as:
\begin{align}
    \langle W_{q}(C)\rangle &= {1\over Z}
    \sum_{ \substack{n_1,n_2  \\  \overline{n}_1,\overline{n}_2} }
    {\calA^{n_1+\overline{n}_1}\calA_{\rm c}^{n_2+\overline{n}_2}\over n_1!\, n_2!\, \overline{n}_1!\, \overline{n}_2!} \left(K\,e^{-S_I}\right)^{n_1+n_2+\overline{n}_1+\overline{n_2}} 
e^{i (\theta + 2 \pi q)  (n_1 -\overline{n}_1) }  \, e^{i  \theta  (n_2-\overline{n}_2)} 
\end{align}
 where  $\calA_{\rm c}$   is the complement of $\calA$ defined above. Performing the sum reproduces  \eqref{VEV}, the area law of confinement.   

\vspace{0.3cm}
In the non-supersymmetric literature, since the $\varrho$ is not introduced, a troubling feature of the dual model  \eqref{dual-AHM} is not addressed. Observe that perturbative plus (leading)  non-perturbative potential in  \eqref{dual-AHM}  is unbounded from below, and as it stands, the lagrangian is troublesome. Usually, when people work with the dilute gas of vortices, one  implicitly assumes that one works around $\varrho=0 $ mean-field. The stability requires addition of a term  $+ e^{- (y + {\bar y})}  = + e^{- 2 \varrho} $ to the effective action. Indeed, such a term do get reproduced in second order in cluster expansion, as well as in the supersymmetric theory, as we explore in the next subsection. 

\subsection{Supersymmetric abelian Higgs model}

In the ${\cal N}=(2,2)$ supersymmetric version  abelian Higgs model with $N=1$, every particle in the spectrum is gapped out by super-Higgs mechanism.   
The vector multiplet $V$  and chiral multiplet $\Phi_1$  acquire  a mass $m= e \sqrt {2r_0}$, leaving no degrees of freedom  in the infrared.

\vspace{0.3cm}
Although the purely bosonic AHM is confining, 
once we introduce a microscopically massless fermion,   we loose confinement for external probe charges. In particular, one looses  the theta  angle dependence of the vacuum energy or put it differently, the external probe charge  
$q$ can be removed by a chiral rotation.   The  ${\cal N}=(2,2)$ theory exhibits perimeter law. 
\begin{align}
W_{q}(C)  \sim e^{- \mu_q   P_C}, \qquad (\rm supersymmetric  \;  case)
\end{align}
where $P_C$ is the perimeter of $C$. 

\vspace{0.3cm}
As such, one may ask the  physical role of  instantons or vortices  in this model. Due to index theorem, the vortex acquires exactly  two-fermionic zero modes.   The form of the vortex operator is: 
\begin{align}
  {\cal V} \sim  e^{-S_v  + i \theta}  e^{- \varrho + i \vartheta  } \; \;   \overline  \chi_{+} \chi_{-} 
  \qquad 
\overline  {\cal V}  \sim e^{-S_v  - i \theta}  e^{  - \varrho (x) -  i  \vartheta (x) - i \theta  }  \;\;  \overline  \chi_{-} \chi_{+}  
\label{vortexSUSYAHM}
\end{align}
Although these vortex operators looks formally same as  $ {\cal V}_1,   \overline {\cal V}_1$ in   \eqref{eq:Vortexsusy},   there is a fundamental difference.   The interactions between the vortices in susy version of $N=1$ abelian Higgs model,   \eqref{vortexSUSYAHM},  are short-ranged and classical interactions are given by \eqref{short-range-susy}. When we calculate their correlators,  we must use propagators for the bosons  and fermions of mass  $m =  e \sqrt {2 r_0}$.  In the infrared, both bosons and fermions freeze even without taking into account non-perturbative contribution. 

\vspace{0.3cm}
The effect of vortices, in the supersymmetric AHM is following. We follow the argument of Ref.~\cite{Hori:2000kt}. 
 The two-point function of the 
fermionic fields $ \chi_{+} (x)  $ and  $\overline  \chi_{-} (y) $  is zero in the free theory of the $Y, \Sigma $ fields. 
\begin{align}
  \langle \chi_{+} (x)  \overline  \chi_{-} (y) \rangle_{\rm free} =0 
  \end{align}
    Yet, non-perturbative effects induced by vortices  \eqref{vortexSUSYAHM} can source such correlation functions.  In the vacuum, we can set the scalars to zero, hence replace $  {\cal V}(z) \rightarrow e^{-S_v  + i \theta}   \overline  \chi_{+} \chi_{-} $. Hence, the effect of one-instanton on the correlator is 
\begin{align}
 \langle \chi_{+} (x)  \overline  \chi_{-} (y) \rangle_{\rm  1-inst. in}  &=   \langle \chi_{+} (x)  \overline  \chi_{-} (y) \int  d^2z \; {\cal V}(z)  \rangle_{\rm  free} \cr
 &=e^{-S_v + i \theta}  \int d^2z \;  S_{++}(x-z)  \;  S_{--}(z-y)  \neq 0 
\end{align}
 where $S_{\pm\pm}$  is the massive Dirac propagator. Crucial point is that this correlator is proportional to $e^{-S_v + i \theta}$ and is due to non-perturbative defects, vortices. 
Later on, this will help us to identify chiral condensate in the  $\CP^{N-1}$ models. 

\vspace{0.3cm}
Hence,   in the supersymmetric theory, vortices   provide a non-perturbative contribution to the superpotential  $(\sim e^{-Y})$  and modifies it into 
\begin{align}
\widetilde{W}=\Sigma\left(Y -t\right)
+\,\mu e^{-Y},
\label{superpot-AHM}
\end{align}
The dual field theory to ${\cal N}=(2,2)$  Abelian Higgs model is described by the free Lagrangian of $Y, \Sigma $  superfields and superpotential  $\widetilde{W}$. 
\begin{align}
L=   \int d^4\theta
\left(\,
-{1\over 2e^2}  \overline \Sigma  \Sigma - \frac{1}{4 r_0}  \overline  Y Y
\,\right)
+{1\over 2}\Biggl(
\int d^2\widetilde{\theta}  \left( \Sigma\left(Y -t\right)
+\,\mu e^{-Y} \right)
\,+\, c.c.\,
\Biggr),
 \label{dual-AHM-susy}
\end{align} 
This is the supersymmetric completion of the non-supersymmetric statistical field theory with Lagrangian \eqref{dual-AHM}. Integrating out gauge field  as in \eqref{atof} produces the mass gap for $\vartheta$, and supersymmetry guarantees that one gets the same mass for all fields in the $Y$-multiplet. The effect of the non-perturbatively induced term is to generate the terms 
$ -e^{-S_v  }  e^{- \varrho + i \vartheta  }   \overline  \chi_{+} \chi_{-}  
-   e^{-S_v  }  e^{  - \varrho (x) -  i  \vartheta (x)   }    \overline  \chi_{-} \chi_{+} + e^{-2 S_v }  e^{-2 \varrho}.$ 
In fact, we can turn on a soft supersymmetry breaking mass term (recall that there is a supersymmetry respecting mass term in the effective Lagrangian as well) of the form 
$m \overline  \chi_{+} \chi_{-}  + \rm {c.c.} $,  and use it to absorb the fermionic  zero modes of the vortex. 
As a result, one lands on an effective potential of the form  
$-   e^{- \varrho } \cos \vartheta   + e^{- 2 \varrho} $ for which the potential is bounded from below.

\subsection{Intermediate \texorpdfstring{$\U(1)^N$}{U(1)N} Abelian Higgs model} 


\vspace{0.3cm}
The $N=1$ supersymmetric $\U(1)$ abelian gauge theory, despite its simplicity,  has an important place in the 
construction of mirror symmetry. The main claim of Ref.\cite{Hori:2000kt} is that the vacuum structure of the $N$ {\it decoupled} copies of the supersymmetric Abelian Higgs model, which is a $\U(1)^{N}$ gauge theory, is continuously connected to the vacuum structure of the  
$\CP^{N-1}$ model.  In the limit where one ungauges   $\U(1)^{N-1}$, the IR theory becomes $\CP^{N-1}$ model. 
One can write the dual Landau--Ginzburg theory for the $\U(1)^{N}$ abelian Higgs model as well. Implementing the same ungauging 
in the dual description, one lands on the dual  Landau--Ginzburg  formulation of the $\CP^{N-1}$ model. We will show that 
our refined instanton analysis is identical to the dual Landau--Ginzburg formulation.

\begin{table}[t]
\begin{center}
\begin{tabular}{| l | l |}
\hline
Susy $\U(1)^{N}$ abelian Higgs model  & Susy $\CP^{N-1}$ \cr 
\hline
$ W_q(C) \sim e^{- \mu_q P(C)}$ & $ W_q(C) \sim e^{- \mu_q P(C)}$   \cr
\hline
$ \langle \chi_{+,i}  \overline  \chi_{-,i}  \rangle  \sim  e^{-S_{v,i} + i \theta_i}  $  & 
$ \langle \psi_{+}  \overline  \psi_{-}  \rangle  \sim e^{-\frac{S_I}{N} + i \frac{\theta}{N} } $
 \cr 
\hline
   gapped  &  gapped \cr
   \hline
mass deformation  $   \Delta {\cal L} \sim m \chi_{+,i}     \overline  \chi_{-,i}  + {\rm c.c}  $
&  mass deformation $  \Delta  {\cal L} \sim m \psi_{+}     \overline  \psi_{-}  + {\rm c.c}   $ \cr 
 $\Longrightarrow  {\rm confinement} \;\;  W_q(C) \sim e^{- {\cal A}(D) T_q}$   & 
 $\Longrightarrow  {\rm confinement} \;\;  W_q(C) \sim e^{- {\cal A}(D) T_q}$ \cr
\hline 
\end{tabular}
\end{center}
\caption{Ungauging   $\U(1)^{N-1}$ part of the abelian Higgs model turns  the theory to $\U(1)$  gauge theory with $N$-chiral field, 
for which the IR-limit is $\CP^{N-1} $ model.  The non-perturbative properties of the   $\U(1)^{N}$ abelian Higgs model are calculable by semi-classical methods. $\CP^{N-1}$ \ model becomes strongly coupled and is incalculable.  }
\label{sym-real}
\end{table}%

 \vspace{0.3cm}
The intermediate  theory that has been used in the construction of the mirror symmetry is a    $\U(1)^N$ gauge theory 
\begin{align}
L= \sum_{i=1}^N \int d^4\theta
\left(\,
 \overline \Phi_i\,e^{2Q_iV_i}\Phi_i
-{1\over 2e^2}  \overline \Sigma_i \Sigma_i
\,\right)
+{1\over 2}\Biggl(
-\int d^2\widetilde{\theta}~t_i\,\Sigma_i
\,+\, c.c.\,
\Biggr),
\label{lagdecoupled}
\end{align} 
which is $N$ decoupled model of the supersymmetric abelian Higgs model.  This theory can be obtained from the GLSM \eqref{lag} 
by gauging the $\U(1)^{N-1}$ subgroup of the global $\SU(N)$ symmetry.  Note that 
 by ungauging  $\U(1)^{N-1}$ part, we should move back to GLSM with full non-abelian $\SU(N)$ global symmetry.

 \vspace{0.3cm}
The   classical potential for \eqref{lagdecoupled} which determines the classical supersymmetric vacua   is given by:
\begin{align}
U= 
 \sum_{i=1}^{N}  |\sigma_i \phi_i|^2  +  {e^2\over 2} \sum_{i=1}^N  \left(\,  |\phi_i|^2-r_{0i}\,\right)^{\!2} 
 \label{pot-vac2}
\end{align}
The classical minima now correspond to: 
\begin{align}
\sigma_i =0, \qquad   |\phi_i|^2=r_{0i}, \qquad i=1, \ldots N
\label{vev2}
\end{align}
 Hence, the vacuum manifold is in fact rather trivial:
 \begin{align}
{\cal M}_{\rm vac} = \Big\{  (\phi_1, \ldots, \phi_N) \Big| |\phi_i|^2= r_{0i} \Big\}/ U(1)^N  = {\rm point}
 \label{cvm2}
\end{align}
 which is  $(\SS^1)^N$  modulo $\U(1)^N$ gauge transformation, 
 therefore, it is a point. At the vacuum, all degrees of freedom are gapped out by the super Higgs mechanism.  Compare 
 \eqref{pot-vac2}  and \eqref{cvm2} with the original U(1) gauge theory vacuum structure  \eqref{pot-vac}  and \eqref{cvm}.

  \vspace{0.3cm}
 The theory dual to $N$ decoupled supersymmetric abelian Higgs model is obviously the sum of  $N$ decoupled  
  Lagrangians  of the form \eqref{dual-AHM-susy}:
 \begin{align}
L=   \sum_i  \int d^4\theta
\Big(
-\textstyle{{1\over 2e^2}}  \overline \Sigma_i  \Sigma_i - \frac{1}{4 r_{0}}  \overline  Y_i Y_i
\,\Big) 
+ \Big(\half
\int d^2\widetilde{\theta}  \Big( \Sigma_i ( Q_i Y_i -t_i)
+\,\mu e^{-Y_i} \Big)
+ {\rm c.c.}
\Big)
 \label{dual-AHM-susy2}
\end{align} 
If we ungauge  $\U(1)^{N-1}$ in \eqref{lagdecoupled}, that theory reduce to gauged linear sigma model with content $(\Phi_i, a)$. For generic $Q_i$,   GLSM has a $\U(1)^{N-1}$ global symmetry. But for $Q_1= \ldots= Q_N$, GLSM has an enhanced $\SU(N)/\ZZ_N$ global symmetry, just like the sigma model. The low energy limit of  GLSM corresponds to taking $e^2 \rightarrow  \infty$,  in which the gauge field becomes non-dynamical, and the theory reduces to the  $\CP^{N-1}$ model.  We have to follow the very same prescription with the dual formulation.

\subsection{Superpotential for the dual of \texorpdfstring{$\CP^{N-1}$}{CPN-1}}

We wish to keep only  $\U(1)_{\rm diag} $ as a  gauge field, and ungauge the remaining $\U(1)^{N-1}$ in the dual theory as well.  This amounts to setting 
$\Sigma_1 = \ldots = \Sigma_N = \Sigma $.  The resulting superpotential is given by  
\begin{align}
\widetilde{W}=\Sigma\left(\sum_{i=1}^NQ_iY_i-t\right)
\,+\,\mu\,\sum_{i=1}^Ne^{-Y_i},
\label{superpot}
\end{align}
where $t= \sum_{i=1}^{N} t_i $. This is the dual of the gauged linear sigma model.  For  $Q_1= \ldots= Q_N$, the original theory has a $\SU(N)$ global symmetry. 
This  symmetry is 
not manifest in the dual formulation, rather only 
\begin{align}
\U(1)^{N-1} \times {\rm Weyl}(\SU(N)) = \U(1)^{N-1} \times S_N 
\end{align} 
where $S_N$ is the permutation group is.   However, this is not a surprise. For example, it is well-known that abelian bosonization maps  $N$-component massive Schwinger model  with  $\SU(N) $ global symmetry to $N$ compact scalars, for which 
manifest global symmetry is again $\U(1)^{N-1} \times S_N$.   The rest of the symmetries are non-linearly realized and not obvious \cite{Coleman:1976uz}.  
 More sophisticated tools such as non-abelian bosonization can make global symmetries manifest.

 \vspace{0.3cm}
The vortex (which are instantons in 2d)  operators in 2d are given by $e^{-Y_i}$.   
 Integrating out  $\Sigma$ at finite $e^2$,  similar to \eqref{atof},  one obtains a mass for  $Y_1+ \ldots +Y_N$ combination. 
This mass is given by $e\sqrt{|r_0|}$ where $r_0$ is bare FI-parameter.  
The modes along the vacuum manifold, which remain massless to all orders in perturbation theory,  acquire masses non-perturbatively, a mass of the order of the strong scale $\Lambda$ of the sigma model.  The precise mechanism of this effect will be described below, 
both by using tools of supersymmetry, and  as well as by using cluster expansion.

 \vspace{0.3cm}
 In the sigma model limit, $ e  \gg \Lambda$, and because of the suppression of the kinetic term,  $   -{1\over 2e^2}  \overline \Sigma\Sigma$, the content of the $\Sigma$ field becomes non-dynamical.  
  As such, we can integrate out  $\Sigma$, producing 
the constraint  $\sum_i Y_i=t$. Equivalently, we can say that the mass term for the $\sum_i Y_i$ mode forces it to its vacuum value, given by the constraint. Either way, one obtains  the superpotential in terms of $N-$1 dynamical $Y_i$ fields: 
 \begin{align}
\widetilde{W}\,=\,\mu\,\sum_{i=1}^Ne^{-Y_i}, 
\end{align}
This superpotential has $N$  minima, obtained by solving 
 \begin{align}
 \partial_{Y_i}  \widetilde{W} =0 \rightarrow 
e^{-Y_{i}^*} =     e^{-t + Y_{1}^* + \ldots+  Y_{N-1}^*} \Longrightarrow  e^{-Y_{i}^*} = e^{-t/N}  e^{i  \frac{2 \pi k}{N}},  \qquad k=1, \ldots, N 
\label{vacua}
\end{align} 
Let  $ \widetilde Y_i$ denote the  fluctuations around the minima, using   $Y_i = \frac{t}{N} +     \widetilde Y_i$. Then, using 
 dimensional transmutation, 
 \begin{align} 
  \mu    e^{-t (\mu) /N}   \equiv    \mu    e^{ - \frac{4 \pi}{g^2(\mu) N }+ i \theta/N}=   \Lambda 
  \end{align}
   where $\Lambda$  is the renormalization group invariant  strong scale of the theory,   we can rewrite 
the superpotential as: 
 \begin{align}
\widetilde{W}\,=\Lambda \;  \sum_{i=1}^Ne^{- \widetilde Y_i}, \qquad  \qquad   \sum_{i=1}^N \widetilde Y_i = 0
\label{suppot2}
\end{align}
  This is the famous result of , from which some of the most interesting aspects of non-perturbative physics follows, such as mass gap and chiral symmetry breaking.   In the bulk of the paper, we rederive this result by our  refined instanton analysis of ${\cal N}=(2,2)$ theory.

\vspace{0.3cm} 
\noindent
{\bf A nice basis for computations:} The superpotential is written in a basis in terms of $N-1$ complex physical fields, $(\tilde y_1,  \tilde y_2, \ldots, \tilde y_{N-1})$, and
we denote $y_N =  -  \sum_{i=1}^{N-1} \tilde y_i $. Recall that   standard irreducible representation of $S_N$
is indeed $N-1$ dimensional. 

Below, we will use a parametrization of $Y$ with one (complex) redundant component, which decouples from the dynamics.  
We  use 
$y = ( y_1, \ldots, y_N) $ fields with no restriction on them. The point is the mode associated with $y_1+  +  y_N $ totally decouples from our description. In our description, we can write 
\begin{align} 
 \widetilde Y_i = \bm \nu_i \cdot Y,  \qquad i= 1, \ldots, N
\end{align}
where a  convenient basis for computations  is given by, 
\begin{align}
 {\bm \nu}_i =  {\bm e}_i - \frac{1}{N} \sum_{j=1}^{N}  {\bm e}_j, \qquad ({\bm e}_i)_j = \delta_{ij},   \qquad  ( {\bm \nu}_i)_j = \delta_{ij}- \frac{1}{N}, \qquad i, j=1, \ldots, N 
\end{align}
We can rewrite the superpotential in terms of these variables as:
 \begin{align}
\widetilde{W} (Y) \,=\, \Lambda \,\sum_{i=1}^N e^{- {\bm \nu}_i. Y }  = \, \Lambda \,  \sum_{i=1}^N e^{- \left(1- \frac{1}{N} \right) Y_i + \frac{1}{N} \sum_{ j \ne j} Y_j   }
\label{suppot3}
\end{align}

\section{Instanton measure} 
In this appendix, we determine the measure over the  instanton collective coordinates in 2d  bosonic $\CP^{N-1}$ model and its generalization with the inclusion of  $n_f$ Dirac fermions. 
The case $n_f=1$ is ${\cal N}=(2,2)$ supersymmetric  theory, and $n_f \geq 2$ are multi-flavor models.  
Instantons in the $n_f$-flavor model  have $2N$ bosonic  and $2Nn_f$ fermionic  zero modes.    
 
The measure of the $k=1$ instanton can be written as a product over the bosonic  and fermionic zero modes as  \begin{align}
 e^{-\frac{4 \pi}{g^2(M_0)}   + i \theta} d\mu_{\rm B}   \;  d\mu_{\rm F} \;.
 \end{align}  
Here, $g^2(M_0)$ is the bare coupling at the Pauli--Villars scale $M_0$.  
The  measure over the bosonic and fermionic coordinates are, respectively,  
\begin{align}
d\mu_{\rm B}  &=    M_{0}^{2N} \;  d^{2N}a  \;    \left(\sqrt {S_0}\right)^{2N}   \frac{1}{ \sqrt{ {\det}'(M_b)} } 
 \cr  
 d\mu_{\rm F} &=    \prod_{a=1}^{n_f}   M_{0}^{-N}   d^{2N}\xi_a \;   (\sqrt {S_0})^{-2N} { {\det}'(M_f) }   
\label{measure3}
\end{align} 
Each bosonic zero mode is associated with a factor of  $M_{0}$ and  $ S_0^{ 1/2}$.    $ [\det' (M_b)]^{- 1/2 }$ denotes determinant over the bosonic   fluctuations around the instanton, where  prime stands for  the removal of zero modes. Similarly, each 
fermionic  zero mode is associated with  $M_{0}^{-  1/2} $ and  $ S_0^{ - 1/2}$, and 
determinant denotes fermionic fluctuations around the instanton.     

The  $2N$  bosonic moduli  parameters of an instanton  are usually decomposed into $x \in \RR^2 $  the center position, 
$\rho \in \RR^{+}$  the size, and $ \Omega \in M_{2N-3}$  the angular moduli,  giving  $ 2 + 1+ (2N-3) =2N $  for its dimension.
In this parametrization, the measure is
\begin{align}
d^{2N}a  = d^2x \;  d\rho \rho^{2N-3}  \; d\Omega  
\end{align}
The fluctuation determinants over the bosonic and fermionic non-zero modes  are related and  can be combined as:
\begin{align}
 [\det (M_b)]^{-1/2}   [ \det (M_f) ]^{n_f} &  =   [\det(-D^2) {\bm 1}_N ]^{n_f-1} 
=   (M_{0} \rho)^{(n_f-1)N}
\end{align} 
The determinants cancel exactly for $n_f=1$ theory, which is the supersymmetric case. 

The explicit $M_0$  dependence in   \eqref{measure3} must cancel by the implicit dependence that enters through the  bare  coupling $g_0= g(M_0)$, and turn into the physical strong scale through dimensional transmutation.   
To see this, let us collect powers of the Pauli--Villars scale $M_{0}$.  
It has contributions from the bosonic and fermionic measures  $M_{0}^{2N - n_f N}$,  and also from the bosonic and fermionic fluctuation operators, 
$ (M_{0})^{(n_f-1)N}$ that combines to give 
\begin{align} 
\underbrace{M_{0}^{2N - n_f N}}_{\rm measure}  \cdot  \underbrace{ M_{0}^{(n_f-1)N}}_{\rm fluc. \; op.} = M_{0}^{N}= M_{0}^{\beta_0}
\end{align} 
where   $\beta_0=N$ is   the  renormalization group beta function coefficient of  the $n_f$ flavor $\CP^{N-1}$ model. 
It is well-known from  standard perturbative calculation  that the one-loop beta functions of the non-linear sigma models is independent of the number of fermionic species.  
This fact also arises from instanton calculus thanks to the cancellation between the factors in the measure and the ones in fluctuation determinant.
The combination gives
\begin{align}
 M_{0}^{\beta_0}  \times e^{- \frac{4 \pi}{g^2(M_0)} } =\Lambda^N  
 \end{align}
where  $\Lambda$  is the renormalization group invariant strong scale of the theory.   
Combining various factors together we obtain: 
  \begin{align}
 &{ \Lambda^N }   \int \prod_{a=1}^{n_f}   d^{2N}\xi_a    \;\;   \Omega_{2N-3} \int {d^2x  d\rho }   
\; \underbrace{ \rho^{2N-3} }_{\rm measure}\;  \underbrace{ \rho^{(n_f-1)N}}_{\rm fluc. \; det. }
 \label{measure5}
\end{align}  

An immediate physical implication of this result is the non-perturbative (topological theta angle dependent) contribution to vacuum energy.  
For the bosonic case,  \eqref{measure5}  reduces to   
   \begin{align}
    \Lambda^N \int  {d^2x  d\rho }   \rho^{N -3},  
   \end{align}
giving rise to the infamous infrared  divergent   contribution to vacuum energy density: 
 \begin{align}
    {\cal E}_{\rm vac} \sim  -   \Lambda^2( e^{ i \theta} +{ \rm c.c.})  \int   d(\Lambda \rho)    (\Lambda \rho)^{N -3} 
\end{align}
as $ \rho \rightarrow \infty$.     
Of course, this is unacceptable.    
Part of our refined instanton analysis is meant to fix this pathological behavior of the standard instanton analysis.

 \section{Fluctuation determinant  for instantons and quantum interactions}

In the bosonic $\CP^{N-1}$ model, the  fluctuation determinant  can be written  by using the two different parametrizations we described in the paper. 
One is to express it as a function of the size modulus $\rho$ and the other is to use the coordinates  $a_1, \ldots, a_N$ for the constituent locations.  
By using the mapping between these two \eqref{mapping0}:
\begin{align}
 [\det (M_b)]^{-1/2}  
&=   (M_{0} \rho)^{-N}   \cr 
&=    M_{0}^{-N}   \Big( { \textstyle \frac{1}{N} }  \sum_{i} |a_i -a_{\rm c} |^2  \Big)^{-\frac{N}{2}}   \cr 
& =    M_{0}^{-N}   \Big(  {\textstyle \frac{1}{2N^2} }   \sum_{i,  j} |a_i -a_j|^2  \Big)^{-\frac{N}{2}}   
\label{det-tworep}  
\end{align}  

It is also possible to reinterpret the effect of the fluctuation determinants in two complementary way. 
Recall that  the  classical interaction  between the fictitious constituent in our construction is zero, $V_{\rm int, cl.} =0$, consistent with the existence of the classical moduli space.  
However, quantum mechanically, the moduli space may be lifted. 
Indeed, the determinant can be rewritten as  $e^{-  V_{\rm int, 1-loop} }$ where $V_{\rm int, 1-loop}$ is 1-loop induced potential.  
One can write the instanton measure in the bosonic theory ($n_f=0$ in \eqref{measure5}) as:
 \begin{align}
 & \Lambda^N   \;     \int   \prod_{i=1}^{N}  {d^2 a_i}  \;     e^{-  V_{\rm int, 1-loop}}   \qquad {\rm where} \cr
    V_{\rm int, 1-loop} &=  \frac{N}{2} \log  \left[  \sum_{i=1}^N |a_i -a_{\rm c}|^2 \right]  +    c_1  \cr
    &=  \frac{N}{2} \log  \left[  \sum_{i,  j=1}^{N} |a_i -a_j|^2 \right]  + c_2 
\label{measure8c}
\end{align} 
where $c_{1} =   \frac{N}{2} \log N $ and $c_{2} =   \frac{N}{2} \log (2N^2) $   are $N$-dependent constants.  

In the theories with fermions, the nature of the quantum-induced interactions between constituents change.   
The net effect of introducing fermions on the 1-loop potential is
  \begin{align}
  V_{\rm int, 1-loop}^{n_f } &=    (1-n_f)    V_{\rm int, 1-loop}. 
 \label{measure10}
\end{align} 
As expected, the one-loop potential vanishes for the supersymmetric theory. 
This allows us safely proceed using just classical interactions between constituents.  
Furthermore, the sign of the quantum-induced interactions is flipped compared to that in the bosonic case in the multi-flavor theories.  
Note that this is in contrast to the classical interactions between constituents \eqref{eq:constituent-interaction}, which are independent of the number of fermion flavors $n_f$.   

In the interaction potential \eqref{measure8c},  the sums  over constituent labels are inside the logarithms, not outside.  
This tells us that the loop-induced interaction between the parameters of the moduli, for $N > 2$, cannot be written as a sum of two-body interactions. 
It is a genuine $N$-body interaction.  
Only for $N=2$ can the one-loop potential be written as a two-body potential:
   \begin{align}
  V_{\rm int, 1-loop} &=  (2-2n_f) \log |a^1-a^2|.
 \label{potential-qu}
\end{align} 

Some remarks are in order.

First, note that the classical interactions   between constituents are robust and are two-body interactions, while the nature of the quantum one-loop interactions changes with the number of fermionic flavors. Indeed, for $N > 2$, the quantum interactions are genuine $N$-body interactions not reducible to two-body interactions.

Another observation is that  \cite{Berg:1979uq} starts with  vacuum energy density  ${\cal E}_{\rm vac} \sim   \int   d \rho    \rho^{N -3} $  which derives from one-loop fluctuation determinant  $[\det (M_b)]^{-1/2}  =   (M_{0} \rho)^{-N} $  (their equation (1)). 
The fluctuation determinant can also be viewed as a one-loop potential for $\rho$ moduli.  
It is  $e^{-V_{\rm int}} =  e^{-N \log(M_0 \rho)}$.  Ref. \cite{Berg:1979uq} ends, after tedious calculation, with a fluctuation determinant for a general instanton in terms of parameters  $c_{\alpha}, a_{\alpha}, \alpha=1, \ldots N$  (their equation (94)). 
But in fact,  $c_{\alpha}$ there are not moduli but encodes the $v$ vector which correspond to the boundary condition of the field at spacetime infinity, $|z| \rightarrow \infty$, and in particular, one does not integrate over them. 
In our construction, we set $c_{\alpha} = 1$ in order to make invariance under the permutation group $S_N$ (which is the Weyl group of   $\SU(N)$) global symmetry manifest.  
With this,  eq.(94) for the one-loop potential in  \cite{Berg:1979uq} reduces to our  \eqref{measure8c}. 
In other words, the simple formula  \cite{Berg:1979uq} starts with (their equation (1)) and the complicated-looking expression it ends with (their equation (94)),  are  actually the same expressions in two different parametrizations of the moduli space, related via the mapping \eqref{mapping0}.

As a final remark, we note that although the fluctuation determinant (and hence one-loop potential) evaluated in \cite{Fateev:1979dc} and  \cite{Berg:1979uq} is correct, it is incorrect to built an effective field theory just based on it. 
Refs.\cite{Fateev:1979dc} and  \cite{Berg:1979uq} argue to do so because the \emph{classical} interactions between instantons (and hence instanton constituents) are zero and the leading interactions seem to be  \emph{quantum}, starting at one-loop level. 
However, the classical Coulomb interaction between  constituents of ${\cal I}$ and $\overline {\cal I}$ are non-zero, and these interactions  ($O(1/g^2)$)  are parametrically  stronger than the quantum ones ($O(g^0)$). 
One can in principle just describe a gas of instantons with quantum interactions, but this has nothing to do with the $\CP^{N-1}$ model. 
As we showed explicitly, and as the mirror symmetry construction of the Landau-Ginzburg dual formulation shows, the statistical field theory is based on the classical interactions between the constituents of ${\cal I}$ and $\overline {\cal I}$.

\bibliographystyle{utphys}
\bibliography{./QFT-Mithat.bib}
\end{document}